\documentclass{article}
\usepackage{amsfonts,amsthm,amssymb}
\usepackage{latexsym,epsf}

\newcommand{\eq}{\begin{equation}}
\newcommand{\en}{\end{equation}}
\newcommand{\eqa}{\begin{eqnarray}}
\newcommand{\ena}{\end{eqnarray}}

\begin{document}

\setlength{\unitlength}{1mm}

\thispagestyle{empty}

 \begin{center}
  { \bf Braid Group, Temperley--Lieb Algebra, and Quantum
     Information and Computation
  \\[2mm]}

  \vspace{.5cm}

  Yong Zhang
 \\[.2cm]

 Department of Physics, University of Utah \\
  115 S, 1400 E, Room 201, Salt Lake City, UT 84112-0830
\\[0.1cm]

\end{center}

\vspace{0.2cm}

\begin{center}
\parbox{13cm}{
\centerline{\small  \bf Abstract}  \noindent\\

In this paper, we explore algebraic structures and low dimensional
topology underlying quantum information and computation. We revisit
quantum teleportation from the perspective of the braid group, the
symmetric group and the virtual braid group, and propose the braid
teleportation, the teleportation swapping and the virtual braid
teleportation, respectively. Besides, we present a physical
interpretation for the braid teleportation and explain it as a sort
of crossed measurement. On the other hand, we propose the extended
Temperley--Lieb diagrammatical approach to various topics including
quantum teleportation, entanglement swapping, universal quantum
computation, quantum information flow, and etc. The extended
Temperley--Lieb diagrammatical rules are devised to present a
diagrammatical representation for the extended Temperley--Lieb
category which is the collection of all the Temperley--Lieb algebras
with local unitary transformations. In this approach, various
descriptions of quantum teleportation are unified in a
diagrammatical sense, universal quantum computation is performed
with the help of topological-like features, and quantum information
flow is recast in a correct formulation. In other words, we propose
the extended Temperley--Lieb category as a mathematical framework to
describe quantum information and computation involving maximally
entangled states and local unitary transformations.

}

\end{center}

\vspace{.5cm}

\begin{tabbing}

\end{tabbing}


\newpage

\section{Introduction}

Quantum entanglements \cite{werner89} play key roles in quantum
information and computation \cite{nielsen99,mermin07} and are widely
exploited in quantum algorithms \cite{shor94, grover97}, quantum
cryptography \cite{bb84, ekert91} and quantum teleportation
\cite{bbcjpw93, bdms00}. On the other hand, topological
entanglements \cite{kauffman02} denote topological configurations
like links or knots which are closures of braids. Aravind
\cite{aravind97} observed that there are natural similarities
between quantum entanglements and topological entanglements. As a
unitary braid has the power of detecting knots or links, it can
often transform a separate quantum state into an entangled one.
Kauffman and Lomonaco \cite{dye02, kl04a} identified a nontrivial
unitary braid representation with a universal quantum gate
\cite{bb02}. Recently, a series of papers have been published on the
application of the braid group \cite{kauffman02} (or the
Yang--Baxter equation \cite{yang67, baxter72}) to quantum
information and computation, see \cite{kl04a, zkg05a, zkg05b} for
unitary solutions of the Yang--Baxter equation as universal quantum
gates; see \cite{kauffman02a, kauffman02b, kl04b} for quantum
topology and quantum computation; see \cite{kl04c, kl03a} for
quantum entanglements and topological entanglements; see for quantum
algebras associated with maximally entangled states
\cite{zjg06,frw06,zg07}; see \cite{zrwwg07,zhang08} for quantum
error correction, topological quantum computing and a possible link
between them.

We focus on the project of setting up a bridge between low
dimensional topology and quantum information, namely looking for low
dimensional topology underlying quantum information and computation.
Kauffman's observation on the teleportation topology \cite{kl04a,
kauffman05} motivates our tour of revisiting in a diagrammatical
approach all tight teleportation and dense coding schemes in
Werner's work \cite{werner01}. In a joint article with Kauffman and
Werner \cite{zkw05}, we make a survey of diagrammatical tensor
calculus and matrix representations, and explore topological and
algebraic structures underlying multipartite entanglements. In this
article as a further extension of our work \cite{zhang06a,zhang06b,
zk07}, we describe quantum information and computation (especially
quantum teleportation \cite{bbcjpw93, bdms00}) in the language of
the braid group and Temperley--Lieb (TL) algebra \cite{tl71}. We
propose {\em the braid teleportation, teleportation swapping and
virtual braid teleportation}, and devise {\em the extended TL
diagrammatical rules} to describe quantum teleportation,
entanglement swapping, universal quantum computation and quantum
information flow.

Quantum teleportation is a procedure of sending a message from
Charlie to Bob with the help of Alice. She shares a maximally
entangled state (for example, Bell states \cite{bell64}) with Bob,
and performs an entangling measurement on the composite system
between Charlie and Alice. After Bob gets results of Alice's
measurement, he is able to obtain the message by exploiting the
protocol between Alice and him. The transformation matrix between
Bell states and product basis is found out to form a unitary braid
representation \cite{kl04a}, and it inspires us to  reformulate {\em
the teleportation equation} (which catches main features
 of quantum teleportation and is defined in the next section)
 in terms of the unitary braid representation $b$-matrix,
 and suggest {\em the braid teleportation}
 $(b^{-1}\otimes Id)(Id\otimes b)$ with identity $Id$ to describe
 quantum teleportation. We present an interpretation for the braid
 teleportation in view of the
 crossed measurement \cite{aav86, vaidman94, vaidman03}, and
 explore the configuration for the
 braid teleportation in the state model \cite{kauffman02} (which
 is devised for the braid representation of the TL
 algebra). Furthermore, the virtual braid group \cite{kauffman99} is
 an extension of the braid group by the symmetric group, and it
 has virtual crossings acting like permutation $P$. We suggest
 {\em the teleportation swapping} $(P\otimes Id)(Id\otimes P)$ as a
 special example of the braid teleportation.
 The virtual mixed relation for defining the virtual braid group is
 found to be a  reformulation of the teleportation equation,
 which leads to our suggestion of {\em the virtual braid teleportation}.
 Moreover, similar to the braid representation of the TL algebra
 \cite{kauffman02}, the virtual braid representation can be
 constructed in terms of the Brauer algebra \cite{brauer37} (or the
 virtual TL algebra \cite{zkw05,zkg06}). The Temperley--Lieb
 configuration for quantum teleportation can be recognized as a
 fundamental configuration defining the diagrammatical Brauer algebra.

 The maximally bipartite entangled pure state is a projector, and
 is able to form a representation of the TL algebra. Based on the
 diagrammatical representation for the TL algebra (i.e., configurations
 in terms of cups and caps \cite{kauffman02}), we devise {\em the
 extended TL diagrammatical rules} to explore topological-like features
 in quantum circuits (or quantum information protocols) involving
 maximally bipartite entangled states and local unitary
 transformations, for examples,
 quantum teleportation, entanglement swapping, universal
 quantum computation, quantum information flow, and etc.
 A maximally entangled Dirac ket (bra) is represented by a
 configuration of a cup (cap), and a local unitary transformation
(its adjoint) is denoted by a solid point (a small circle).
 In our extended TL diagrammatical framework, various approaches to
 quantum teleportation have a unified diagrammatical description, and
 they include its standard description \cite{bbcjpw93, bdms00},
 the transfer operator (or quantum information flow) \cite{preskill},
 measurement-based quantum teleportation \cite{vaidman94}, and
 Werner's tight teleportation schemes \cite{werner01}. The transfer
 operator is described by a configuration involving a top cap and a
 bottom cup in which the teleportation appears to be a kind of the
 flow of quantum information. The measurement-based quantum
 teleportation has a typical TL configuration as a
 product of generators of the TL algebra, and this diagram
 is able to describe both discrete and continuous quantum
 teleportation schemes. The diagram describing the tight
 teleportation scheme is a closure of the configuration for
 measurement-based quantum teleportation, and it naturally derives
 a characteristic equation for quantum teleportation since a closed
 configuration in the extended TL diagrammatical rules corresponds
 to a trace of products of operators.

 The extended TL diagrammatical rules present a diagrammatical
 representation of {\em the extended TL algebra} as an extension of
 the TL algebra by local unitary transformations. The collection of
 all the extended TL algebras is
 called {\em the extended TL category} \cite{zhang06a,zhang06b,zk07}, and its
 diagrammatical representation includes all configurations made of
 cups, caps, solid points and small circles. Besides its
 application to quantum teleportation, it is able to describe
 entanglement swapping \cite{ekert93}, universal quantum computation
 \cite{gc99}, and etc. Entanglement swapping is an approach to producing
 an entangled state between two independent systems via quantum measurements,
 and the closure of its diagrammatical description gives rise to the tight
 entanglement swapping scheme with a characteristic equation.
 Universal quantum computation is performed in the extended TL
 category, since we are able to construct unitary
 braid gates, the swap gate and CNOT gate, etc., with the extended
 TL diagrammatical rules. Furthermore, we recognize another
 equivalent description of quantum teleportation in terms of the
 swap gate and Bell measurements, after identifying the
 configuration  for quantum teleportation with that for
 the axiom of the Brauer algebra \cite{zkw05,zkg06}. Moreover, we
 comment on multipartite entanglements in the extended TL diagrammatical
 approach.

  Quantum teleportation can be viewed as a flow of quantum information
  from the sender to the receiver, and hence quantum information
  flow can be well described in the extended TL diagrammatical
  framework. In its configuration, the flow is only a part of the entire
  diagram, related to or even controlled by other parts. For example,
  it is zero due
  to the vanishing trace of a product of local unitary transformations
  which are not involved in the flow. Besides our diagrammatical
  approach, quantum information flow has been described by
  Kauffman's teleportation topology \cite{kl04a, kauffman05} and
  Abramsky and Coecke's strongly compact closed categories \cite{coecke04}.
  There are essential physical and mathematical differences among
  them. Measurement-based quantum teleportation is chosen to present
  a full description of quantum teleportation in our study, whereas
  quantum information flow denoted by the transfer operator is
  regarded as the entire quantum teleportation in both
  \cite{kl04a, kauffman05} and  \cite{coecke04}. We show that
  the paradigm described by the transfer operator is a part of the
  picture by measurement-based quantum teleportation.
  Furthermore, only topological-like features
  \cite{zk07} can be explored in the extended TL diagrammatical configuration
  instead of pure topology in \cite{kl04a, kauffman05}. Moreover, we
  propose the extended TL category underlying quantum information
  protocols like quantum teleportation instead of
  strongly compact closed categories, see \cite{zk07} for more
  details.

The plan of this paper is organized as follows. Section 2 introduces
the teleportation equation and reformulates it respectively by the
braid group, the symmetric group and the virtual braid group.
Section 3-7 introduces the extended TL diagrammatical approach to
quantum information and computation: Section 3 explains
diagrammatical rules with examples; Section 4 unifies various
descriptions of quantum teleportation at the diagrammatical level;
Section 5 focuses on the TL algebra and the Brauer algebra; Section
6 deals with the entanglement swapping and universal quantum
computation; Section 7 compares the quantum information flow in the
extended TL category with other known approaches. Last section
comments on our work in the project of setting up categorical
foundations for quantum physics and information.

 \section{Braid teleportation, teleportation swapping
   and virtual braid teleportation}

We describe quantum teleportation in the language of the braid
group, the symmetric group, and the virtual braid group,
respectively,  and  propose the braid teleportation, the
teleportation swapping, and the virtual braid teleportation. First
of all, we revisit the standard description of the teleportation
\cite{bbcjpw93, bdms00} and assign the name {\em the teleportation
equation} to its most important equality. Secondly, we reformulate
the teleportation equation in terms of the Bell matrix which forms a
unitary braid representation, and then realize that a braiding
operator called {\em the braid teleportation} plays a key role in
the formulation of the teleportation equation. Thirdly, we discuss
{\em the teleportation swapping} as a simplest example of the braid
teleportation, and recognize the teleportation equation as a
reformulation of the virtual mixed relation defining the virtual
braid group. Lastly, we look upon the braid teleportation as a kind
of crossed measurement if it has a sort of physical correspondence,
and expand it in the state model \cite{kauffman02} which sheds an
insight on the main topic in the following sections.

\subsection{Quantum teleportation: the teleportation equation}

The Pauli matrices $\sigma_1$, $\sigma_2$ and $\sigma_3$ have the
conventional form, \eq \sigma_1=\left(\begin{array}{cc}
0 & 1 \\
1 & 0 \end{array}\right), \qquad \sigma_2=\left(\begin{array}{cc}
0 & -i \\
i & 0 \end{array}\right), \qquad \sigma_3=\left(\begin{array}{cc}
1 & 0 \\
0 & -1 \end{array}\right), \en and quantum states $|0\rangle$ and
$|1\rangle$ as a basis denoting a qubit have the coordinate
presentation in the complex field $\mathbb{C}$,
 \eq
 |0\rangle=\left(\begin{array}{c}
  1 \\ 0 \end{array}\right), \qquad |1\rangle=\left(\begin{array}{c}
  0 \\ 1 \end{array}\right),
  \en
which give rise to useful formulas: $a,b\in {\mathbb C}$, \eq
 \sigma_1 \left(\begin{array}{c}
  a \\ b \end{array}\right)=\left(\begin{array}{c}
  b \\ a \end{array}\right), \,\,\, -i \sigma_2 \left(\begin{array}{c}
  a \\ b \end{array}\right)=\left(\begin{array}{c}
 - b \\ a \end{array}\right), \,\,\, \sigma_3 \left(\begin{array}{c}
  a \\ b \end{array}\right)=\left(\begin{array}{c}
  a \\ -b \end{array}\right).
\en  The product basis of two-fold tensor products denoting
two-qubit
 is chosen to be
 \eq
|00\rangle=\left(\begin{array}{c}
  1 \\ 0 \\ 0 \\ 0 \end{array}\right), \,\,\,
 |01\rangle=\left(\begin{array}{c}
  0 \\ 1  \\0 \\ 0 \end{array}\right), \,\,\,
 |10\rangle=\left(\begin{array}{c}
  0 \\ 0  \\1 \\ 0 \end{array}\right), \,\,\,
 |11\rangle=\left(\begin{array}{c}
  0 \\ 0  \\0 \\ 1 \end{array}\right)
 \en
 which fixes our rule for calculating the tensor product of
 matrices, i.e., embedding the right matrix into the left one.
With the product basis $|ij\rangle$, $i,j=0,1$, four mutually
orthogonal Bell states  have the form, \eqa
  & & |\phi^+\rangle=\frac 1 {\sqrt 2} (|00\rangle+|11\rangle),
  \qquad |\phi^-\rangle=\frac 1 {\sqrt 2} (|00\rangle-|11\rangle),
  \nonumber\\
  & & |\psi^+\rangle=\frac 1 {\sqrt 2} (|01\rangle+|10\rangle),
  \qquad |\psi^-\rangle=\frac 1 {\sqrt 2} (|01\rangle-|10\rangle),
\ena which derive the product basis $|ij\rangle$ in terms of  Bell
states, \eqa
  & & |00\rangle =\frac 1 {\sqrt 2} (|\phi^+\rangle
  +|\phi^-\rangle),\qquad |01\rangle =\frac 1 {\sqrt 2} (|\psi^+\rangle
  +|\psi^-\rangle), \nonumber\\
  & & |10\rangle =\frac 1 {\sqrt 2} (|\psi^+\rangle
  -|\psi^-\rangle),\qquad |11\rangle =\frac 1 {\sqrt 2} (|\phi^+\rangle
  -|\phi^-\rangle).
\ena
 Bell states can be transformed to each other with local unitary
 transformations consisting of Pauli matrices and identity matrix $1\!\! 1_2$,
 \eqa
 \label{local}
 & & |\phi^{-} \rangle=(1\!\! 1_2\otimes \sigma_3) |\phi^+\rangle =
  (\sigma_3 \otimes 1\!\! 1_2 ) |\phi^+\rangle, \nonumber\\
 &&  |\psi^+\rangle =(1\!\! 1_2\otimes \sigma_1) |\phi^+\rangle
  =(\sigma_1 \otimes 1\!\! 1_2) |\phi^+\rangle,
 \nonumber\\
 & & |\psi^-\rangle =( 1\!\! 1_2\otimes -i \sigma_2) |\phi^+\rangle
  =(i \sigma_2\otimes 1\!\! 1_2) |\phi^+\rangle,
 \ena
where one-qubit unitary transformations are called local unitary
transformations.

\begin{figure}
\begin{center}
\epsfxsize=10.cm \epsffile{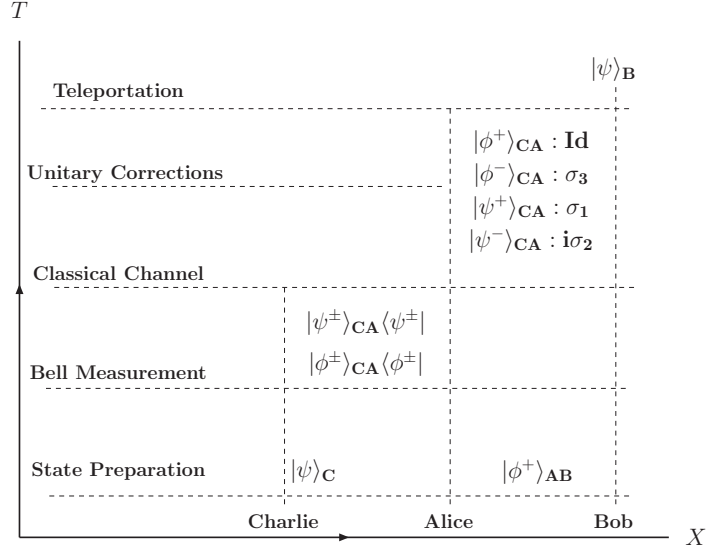}
\caption{Diagrammatical description of quantum teleportation.}
\label{fig0}
\end{center}
\end{figure}

Quantum teleportation  transports a unknown quantum state
$|\psi\rangle_C=$$(a|0\rangle + b |1\rangle)_C$ from the sender,
Charlie to the receiver, Bob, with the help of Alice, and it
exploits properties of quantum entanglement and quantum measurement.
Figure 1 is our diagrammatical interpretation for quantum
teleportation. Let Alice and Bob share the Bell state
$|\phi^+\rangle_{AB}$, a maximally bipartite entangled pure state.
Do calculation:  \eqa \label{tele}
  & |\psi\rangle_C|\phi^+\rangle_{AB} = \frac 1 {\sqrt 2}
 ( a|0\rangle + b |1\rangle)_C (|00\rangle
 +|11\rangle)_{AB}\nonumber\\
 &= \frac 1 2 a (|\phi^+\rangle + |\phi^-\rangle)_{CA} |0\rangle_B
  + \frac 1 2 a (|\psi^+\rangle + |\psi^-\rangle)_{CA} |1\rangle_B
  \nonumber\\
  &  + \frac 1 2 b (|\psi^+\rangle - |\psi^-\rangle)_{CA} |0\rangle_B
  + \frac 1 2 b (|\phi^+\rangle - |\phi^-\rangle)_{CA} |1\rangle_B
  \nonumber\\
  &= \frac 1 2 ( |\phi^+\rangle_{CA}|\psi\rangle_B +
     |\phi^-\rangle_{CA} \sigma_3|\psi\rangle_B
     +  |\psi^+\rangle_{CA} \sigma_1|\psi\rangle_B
     +|\psi^-\rangle_{CA} (-i\sigma_2)|\psi\rangle_B )
\ena which is an identity but called {\em the teleportation
equation} in our research for convenience. Alice performs Bell
measurements in the composite system of Charlie and her, and obtains
four kinds of outcomes. After Alice detects the Bell state
$|\phi^+\rangle_{CA}$ and informs it to Bob through a classical
channel, Bob knows that he has the quantum state $|\psi\rangle_B$
which Charlie wants to send to him. As Alice gets Bell states
$|\phi^-\rangle_{CA}$ or $|\psi^+\rangle_{CA}$ or
$|\psi^-\rangle_{CA}$, Bob applies local unitary transformations
$\sigma_3$ or $\sigma_1$ or $i\sigma_2$  respectively, on the
quantum state that he has to obtain $|\psi\rangle_B$.

There exist beautiful mathematical structures underlying quantum
teleportation, though only fundamental laws of quantum mechanics and
a little linear algebra are involved in its standard description. In
this paper, we make it clear that quantum teleportation can be
described by the braid group, the symmetric group, the virtual braid
group, the TL algebra and the Brauer algebra.

 \subsection{Teleportation equation in terms of Bell matrix}

Let us introduce the Bell matrix \cite{kl04a, zkg05a,dye02}, denoted
by $B=(B_{ij,\,lm})$, $i,j,l,m=0,1$. As a unitary transformation
matrix from the product basis to the basis formed by Bell states, it
forms a unitary braid representation as well as a universal quantum
gate in universal quantum computation. The $B$ matrix and
 its inverse $B^{-1}$ (or transpose $B^T$) have the form  \eq \label{bell}
 B=\frac 1 {\sqrt 2}\left(
 \begin{array}{cccc}
 1 & 0 & 0 & 1 \\
 0 & 1 & -1 & 0 \\
 0 & 1 & 1 & 0 \\
 -1 & 0 & 0 & 1 \\
 \end{array} \right), \qquad  B^{-1}=B^{T}=\frac 1 {\sqrt 2}\left(
 \begin{array}{cccc}
 1 & 0 & 0 & -1 \\
 0 & 1 & 1 & 0 \\
 0 & -1 & 1 & 0 \\
 1 & 0 & 0 & 1 \\
 \end{array} \right).
\en It has an exponential formalism given by \eq
 B=e^{i\frac {\pi} 4 (\sigma_1\otimes \sigma_2)}=\cos{\frac {\pi}
 4}+ i \sin \frac {\pi} 4 (\sigma_1\otimes \sigma_2)
\en with interesting properties: \eq B^2=i\sigma_1\otimes \sigma_2,
\qquad B^4=-1\!\! 1_4, \qquad B^8=1\!\! 1_4, \qquad
 B=\frac 1 {\sqrt 2} (1\!\! 1_4 + B^2). \en
In terms of the $B$ matrix and product basis $|ij\rangle$, Bell
states is yielded in two ways: \eqa
 (I): & & |\phi^+\rangle=B|11\rangle, \qquad  |\phi^-\rangle=B|00\rangle,
 \nonumber\\
  & & |\psi^+\rangle=B|01\rangle, \qquad  |\psi^-\rangle=-B|10\rangle,
 \ena
and
 \eqa
(II): & & |\phi^+\rangle=B^T|00\rangle, \qquad
|\phi^-\rangle=-B^T|11\rangle,
 \nonumber\\
  & & |\psi^+\rangle=B^T|10\rangle, \qquad  |\psi^-\rangle=B^T|01\rangle,
 \ena
where the Bell operator $B$ acts on the basis $|ij\rangle$ in the
way \eq
 B|ij\rangle =\sum_{k,l=0}^{1} |kl\rangle B_{kl,ij}=
 \sum_{k,l=0}^{1} |kl\rangle B^T_{ij,kl}.
\en

For simplicity, we exploit the first type of transformation law
$(I)$ between Bell states and the product basis,  and rewrite the
teleportation equation (\ref{tele}) into a new formalism,
 \eqa
 & & (1\!\! 1_2 \otimes B)(|\psi\rangle\otimes |11\rangle)_{CAB}
 \nonumber\\
 & & =
 \frac 1 2 ( B\otimes 1\!\! 1_2 ) (|00\rangle\otimes \sigma_3 |\psi\rangle
 + |01\rangle\otimes \sigma_1 |\psi\rangle
 + |10\rangle\otimes i\sigma_2 |\psi\rangle
 +  |11\rangle\otimes  |\psi\rangle )_{CAB}, \nonumber\\
 & & \equiv (B\otimes 1\!\! 1_2)({\vec v}^T \otimes \frac 1 2 {\vec\sigma}_{11} |\psi\rangle
 )_{CAB}
 \ena
in which the vector $\vec{v}$, its transpose given by \eq
 {\vec v}^T=(|00\rangle,|01\rangle,|10\rangle,|11\rangle), \qquad
 v_{ij}=|ij\rangle, \,\, i, j=0,1,
\en the vector $\vec{\sigma}_{11}$, its transpose given by \eq
\vec{\sigma}_{11}^{T}=(\sigma_3, \sigma_1, i\sigma_2, 1\!\! 1_2),
\en and the calculation of ${\vec v}^T \otimes {\vec\sigma}_{11}$
follows the rule, \eq {\vec v}^T \otimes {\vec\sigma}_{11}
|\psi\rangle \equiv |00\rangle\otimes \sigma_3 |\psi\rangle
 + |01\rangle\otimes \sigma_1 |\psi\rangle
 + |10\rangle\otimes i\sigma_2 |\psi\rangle
 +  |11\rangle\otimes  |\psi\rangle.   \en

The remaining three teleportation equations are derived in a similar
way with the help of local unitary transformations among Bell states
(\ref{local}). For example,
 \eqa
 & & |\psi\rangle_C|\phi^-\rangle_{AB}= (1\!\! 1_2 \otimes B)
  (|\psi\rangle\otimes |00\rangle)_{CAB}
 \nonumber\\
 &=&|\psi\rangle_C\otimes (1\!\! 1_2\otimes \sigma_3)|\phi^+\rangle_{AB}
 =(1\!\! 1_2\otimes 1\!\! 1_2\otimes \sigma_3)|\psi\rangle_C|\phi^+\rangle_{AB}
 \nonumber\\
 &=& (B\otimes 1\!\! 1_2)({\vec v}^T \otimes \frac 1 2 \sigma_3{\vec\sigma}_{11}
 |\psi\rangle)_{CAB}
 \ena
where the local unitary transformation $1\!\! 1_2\otimes1\!\!
1_2\otimes \sigma_3$ commutes with $B\otimes 1\!\! 1_2$, and  the
other two teleportation equations have the form, \eqa & & (1\!\! 1_2
\otimes B)(|\psi\rangle\otimes |01\rangle)_{CAB}=(B\otimes 1\!\!
1_2)({\vec v}^T \otimes \frac 1 2 \sigma_1{\vec\sigma}_{11}
 |\psi\rangle)_{CAB}, \nonumber\\
 & &(1\!\! 1_2 \otimes B)(|\psi\rangle\otimes
-|10\rangle)_{CAB}=(B\otimes 1\!\! 1_2)({\vec v}^T \otimes -\frac 1
2 i \sigma_2{\vec\sigma}_{11} |\psi\rangle
 )_{CAB}.
\ena These four teleportation equations can be collected into an
equation,
 \eqa \label{btele}
 & & (1\!\!1_2 \otimes B)(|\psi\rangle\otimes {\vec v}^T)_{CAB}
 =(B\otimes 1\!\!1_2) ( {\vec v}^T \otimes \frac 1 2 \vec{\Sigma}|\psi\rangle)_{CAB},
 \nonumber\\
 & &
  ( {\vec v}^T \otimes \frac 1 2 \vec{\Sigma}|\psi\rangle)_{CAB}
  =(B^{-1}\otimes 1\!\!1_2)(1\!\!1_2 \otimes B)(|\psi\rangle\otimes {\vec
  v}^T)_{CAB},
 \ena
 in terms of  the new matrix $\vec{\Sigma}$, a convenient notation given by
    \eq \vec{\Sigma}=(\sigma_3, \sigma_2, i\sigma_1, 1\!\!
 1_2) \vec{\sigma}_{11} \en
where $ {\vec v}^T \otimes \vec{\Sigma}$ is defined as \eq {\vec
v}^T \otimes \vec{\Sigma} |\psi\rangle
 \equiv ({\vec v}^T \otimes \sigma_3{\vec\sigma}_{11} |\psi\rangle,
  {\vec v}^T \otimes \sigma_2{\vec\sigma}_{11} |\psi\rangle,
  {\vec v}^T \otimes i \sigma_1 {\vec\sigma}_{11} |\psi\rangle,
   {\vec v}^T \otimes {\vec\sigma}_{11} |\psi\rangle), \en
and $|\psi\rangle\otimes {\vec v}^T$ has the form \eq
 |\psi\rangle\otimes {\vec v}^T \equiv (|\psi\rangle\otimes |00\rangle,
   |\psi\rangle\otimes |01\rangle,
  |\psi\rangle\otimes |10\rangle, |\psi\rangle\otimes |11\rangle).
\en

Obviously, the operator $(B^{-1}\otimes 1\!\!1_2)(1\!\!1_2 \otimes
B)$ plays a fundamental role in the new formulations of the
teleportation equation. In the following, we verify the Bell matrix
$B$ a unitary braid representation, and then name the braiding
operator of this kind as {\em the braid teleportation}.

\subsection{Braid teleportation and teleportation swapping}

A braid representation $b$-matrix is a $d \times d $ matrix acting
on $ V\otimes V$ where $V$ is an $d$-dimensional vector space over
the complex field $\mathbb{C}$. The symbol $b_i$ denotes the braid
$b$ acting on the tensor product $V_i\otimes V_{i+1}$. The classical
braid group $B_{n}$ is generated by braids $b_1, b_2$, $\cdots,
b_{n-1}$ satisfying the braid group relation,
  \eqa \label{bgr}
   b_{i} b_{j} &=& b_{j} b_{i}, \qquad  j \neq i \pm 1, \nonumber\\
  b_{i}  b_{i+1} b_{i} &=& b_{i+1} b_{i} b_{i+1},
  \qquad i=1, \cdots, n-2.
 \ena
The virtual braid group $VB_n$ is an extension of the classical
braid group $B_n$ by the symmetric group $S_n$ \cite{kauffman99}. It
has two types of generators:  braids $b_i$ and virtual crossings
$v_i$ defined by the virtual crossing relation, \eqa \label{vbgr1}
 v_i^2 &=& Id, \qquad  v_i v_{i+1} v_i = v_{i+1} v_i v_{i+1},
 \qquad i=1,\cdots, n-2,
 \nonumber\\
 v_i v_j &=& v_j v_i, \qquad j \neq i \pm 1,
\ena a presentation of the symmetric group $S_n$ with identity $Id$,
and they satisfying the virtual mixed relation, \eqa \label{vbgr2}
b_i v_j &=& v_j b_i,
\qquad   j \neq i \pm 1, \nonumber\\
 b_{i+1}v_{i}v_{i+1} &=& v_{i}v_{i+1} b_{i}, \qquad i=1,\cdots, n-2.
\ena

 The Bell matrix $B$ forms a unitary braid representation.
 After some algebra, the right handside of the braid group relation
 (\ref{bgr}) has a form
 \eqa
  & & (1\!\! 1_2\otimes B)(B\otimes 1\!\! 1_2)(1\!\! 1_2\otimes B)
  \nonumber\\
    & & = \frac 1 {2\sqrt{2}} (1\!\! 1_8+1\!\! 1_2\otimes i\sigma_1\otimes \sigma_2 )
   ( 1\!\! 1_8 + i\sigma_1\otimes \sigma_2 \otimes 1\!\! 1_2)
   (1\!\! 1_8+ 1\!\! 1_2 \otimes i\sigma_1 \otimes \sigma_2  )
   \nonumber\\
   & &= \frac i {\sqrt{2}} (1\!\! 1_2\otimes \sigma_1\otimes \sigma_2
      + \sigma_1\otimes \sigma_2 \otimes 1\!\! 1_2)
   = \frac 1 {\sqrt{2}} (1\!\! 1_2\otimes B^2
      + B^2\otimes 1\!\! 1_2)
 \ena
and its left handside leads to the same result,  \eq (B\otimes 1\!\!
1_2) (1\!\! 1_2\otimes B)(B\otimes 1\!\! 1_2)=
 \frac 1 {\sqrt{2}} (1\!\! 1_2\otimes B^2  + B^2\otimes 1\!\! 1_2).
\en Furthermore, the Bell matrix $B$ as a classical crossing and the
permutation matrix $P$ as a virtual crossing,
\eq P=\left(\begin{array}{cccc} 1 & 0 & 0 & 0\\
 0 & 0 & 1 & 0\\
 0 & 1 & 0 & 0 \\
 0 & 0 & 0 & 1
\end{array} \right), \qquad P|ij\rangle=|ji\rangle, \qquad i,j=0,1,
\en form a unitary virtual braid representation. The left handside
of the virtual mixed relation (\ref{vbgr2}) has a form \eq (1\!\!
1_2 \otimes B)(P\otimes 1\!\!
 1_2)(1\!\! 1_2 \otimes P)(|i\rangle \otimes |j\rangle \otimes|k\rangle)
 = (1\!\! 1_2 \otimes B)(|k\rangle\otimes |ij\rangle),
 \en
while its right handside leads to the form
 \eqa
 & & (P\otimes 1\!\! 1_2) (1\!\! 1_2\otimes P)(B\otimes 1\!\! 1_2)
  (|i\rangle \otimes |j\rangle \otimes|k\rangle) \nonumber\\
  & & =\sum_{k,l=0}^1 B_{i^\prime j^\prime,ij}(P\otimes 1\!\! 1_2)
    (1\!\! 1_2\otimes P)
 (|i^\prime j^\prime\rangle \otimes |k\rangle) \nonumber\\
& &= \sum_{k,l=0}^1 B_{i^\prime j^\prime,ij}
 (|k\rangle \otimes |i^\prime j^\prime\rangle) =
 (1\!\! 1_2 \otimes B)(|k\rangle\otimes |ij\rangle).
  \ena

\begin{figure}
\begin{center}
\epsfxsize=12.cm \epsffile{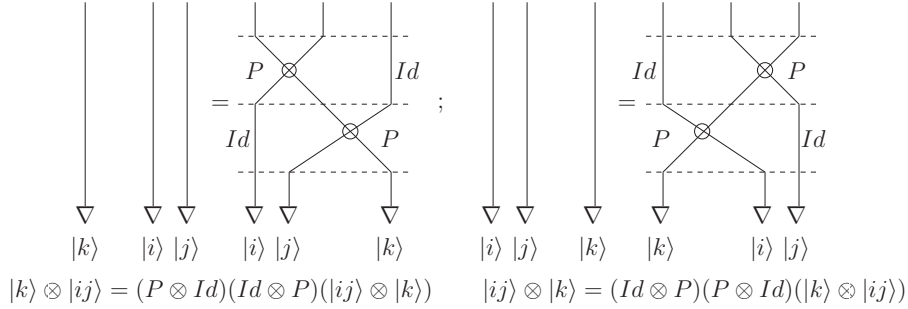} \caption{Teleportation
swapping.} \label{fig1}
\end{center}
\end{figure}

Here we suggest the unitary braiding operator $(b^{-1}\otimes Id)(Id
\otimes b)$ as {\em the braid teleportation} underlying quantum
teleportation. Its simplest example $(P\otimes Id)(Id \otimes P)$ is
called {\em the teleportation swapping} in view of the fact that the
braid group (\ref{bgr}) is a generation of of the symmetric group
(\ref{vbgr1}), and it has the other example $(B^{-1}\otimes
1\!\!1_2)(1\!\!1_2 \otimes B)$ in terms of the Bell matrix. There
are two natural teleportation swapping operators $(P\otimes Id)(Id
\otimes P)$ and $(Id \otimes P)(P\otimes Id)$ in terms of the
permutation operator $P$, satisfying the following teleportation
swapping equalities, \eqa & &
  |k\rangle \otimes |ij\rangle
 =(P\otimes Id)(Id\otimes P) (|ij\rangle \otimes |k\rangle),
 \nonumber\\
 & & |ij\rangle\otimes |k\rangle
 =(Id\otimes P)(P\otimes Id) (|k\rangle \otimes |ij\rangle),
 \ena
which are shown in Figure 2, the virtual crossing $P$ denoted by a
cross with a small circle.

\subsection{Virtual braid teleportation}

Let us describe quantum teleportation in the language of the virtual
braid group. The braid group relation (\ref{bgr}) represents a
connection between topological entanglements and quantum
entanglements, while the virtual mixed relation (\ref{vbgr2}) is a
reformulation of the teleportation equation (\ref{tele}). A
nontrivial unitary braid detecting knots or links can be often
identified with a universal quantum gate transforming a separate
state into an entangled one, see \cite{kl04a, zkg05a, zkg05b}, and
it acts as a device yielding an entangled source. On the other hand,
the teleportation swapping in terms of a virtual crossings $v_i$ is
responsible for quantum teleportation. In the following, the virtual
mixed relation (\ref{vbgr2}) is shown as a reformulation of the
teleportation equation (\ref{tele}).

Note that the permutation matrix $P$ has a form
 \eq
 \label{perm}
 P=\frac 1 2 (1\!\! 1_4 + \sigma_1\otimes \sigma_1
 +\sigma_2\otimes \sigma_2 +\sigma_3 \otimes
 \sigma_3),
 \en
and local unitary transformations (\ref{local}) among Bell states
are given by \eqa
 & & |\phi^{-} \rangle=B|00\rangle= (1\!\! 1_2\otimes \sigma_3) B |11\rangle =
  (\sigma_3 \otimes 1\!\! 1_2 ) B |11\rangle, \nonumber\\
 &&  |\psi^+\rangle = B|01\rangle= (1\!\! 1_2\otimes \sigma_1) B |11\rangle
  =(\sigma_1 \otimes 1\!\! 1_2) B |11\rangle,
 \nonumber\\
 & & |\psi^-\rangle =-B |10\rangle =  ( 1\!\! 1_2\otimes -i \sigma_2) B |11\rangle
  =(i \sigma_2\otimes 1\!\! 1_2) B |11\rangle.
 \ena
In terms of the Bell matrix and teleportation swapping, the left
handside of the teleportation equation (\ref{tele}) has a
 form,
 \eqa
 & &|\psi\rangle_C\otimes |\phi^+\rangle_{AB}=
 (1\!\! 1_2\otimes B)(|\psi\rangle_C\otimes |11\rangle_{AB})
 \nonumber\\
 & & =(1\!\! 1_2 \otimes B)(P\otimes 1\!\! 1_2)(1\!\! 1_2 \otimes P)
 (|11\rangle_{CA}\otimes |\psi\rangle_B),
 \ena
while its right handside leads to the other form,
 \eqa
& & \frac 1 2 (|\phi^{-}\rangle_{CA} \sigma_3 |\psi\rangle_B +
|\psi^{-}\rangle_{CA} (-i\sigma_2) |\psi\rangle_B +
 |\psi^{+}\rangle_{CA} \sigma_1 |\psi\rangle_B + |\phi^+\rangle_{CA}|\psi\rangle_B )
 \nonumber\\
  &=& \frac 1 2 (1\!\! 1_2\otimes \sigma_3\otimes \sigma_3
 + 1\!\! 1_2 \otimes i\sigma_2 \otimes i \sigma_2 + 1\!\! 1_2
 \otimes \sigma_1\otimes \sigma_1 +1\!\! 1_8 ) (B\otimes 1\!\! 1_2)
 (|11\rangle_{CA}\otimes |\psi\rangle_B) \nonumber\\
 &=& (1\!\! 1_2 \otimes P -1\!\! 1_2 \otimes \sigma_2 \otimes \sigma_2)
   (B\otimes 1\!\! 1 _2) (|11\rangle_{CA}\otimes |\psi\rangle_B)
 \ena
Hence the teleportation equation (\ref{tele}) can be  recognized to
be either a kind of the teleportation swapping, \eqa \label{vtele}
 |\psi\rangle_C\otimes|\phi^+\rangle_{AB} &=&(1\!\! 1_2\otimes
 P-1\!\! 1_2\otimes \sigma_2\otimes \sigma_2)(|\phi^+\rangle_{CA}\otimes
|\psi\rangle_{B})\nonumber\\
&=& (P\otimes 1\!\! 1_2)(1\!\! 1_2\otimes P)
(|\phi^+\rangle_{CA}\otimes |\psi\rangle_{B}), \ena or a
reformulation of the virtual mixed relation (\ref{vbgr2}), \eqa
 & & (1\!\! 1_2\otimes B)(P\otimes 1\!\! 1_2)(1\!\! 1_2\otimes P)
 (|11\rangle_{CA}\otimes |\psi\rangle_{B}) \nonumber\\
 & &=(1\!\! 1_2\otimes P-1\!\! 1_2\otimes \sigma_2\otimes \sigma_2)
 (B\otimes 1\!\! 1_2) (|11\rangle_{CA}\otimes |\psi\rangle_{B})
 \nonumber\\
 & &=(P\otimes 1\!\! 1_2)(1\!\! 1_2\otimes P)(B\otimes 1\!\! 1_2)
 (|11\rangle_{CA}\otimes |\psi\rangle_{B}).
\ena

The remaining three teleportation equations can be reformulated by
applying local unitary transformations to the teleportation equation
(\ref{vtele}). The teleportation equation for the Bell state
$|\phi^-\rangle_{AB}$ is obtained to be \eqa && |\psi\rangle_C
\otimes |\phi^-\rangle_{AB}
 =(1\!\! 1_2 \otimes \sigma_3 \otimes 1\!\! 1_2)
 ( |\psi\rangle_C \otimes |\phi^+\rangle_{AB})
 \nonumber\\
 &=& (1\!\! 1_2 \otimes \sigma_3 \otimes 1\!\! 1_2)
 (1\!\! 1_2\otimes P - 1\!\! 1\otimes \sigma_2\otimes
 \sigma_2) (1\!\! 1_2 \otimes \sigma_3 \otimes 1\!\! 1_2)
 (|\phi^-\rangle_{CA} \otimes |\psi\rangle_B) \nonumber\\
  &=& (1\!\! 1_2 \otimes P - 1\!\! 1_2 \otimes \sigma_1\otimes \sigma_1)
   (|\phi^-\rangle_{CA}\otimes |\psi\rangle_B),
 \ena
and the teleportation equations for Bell states
$|\psi^\pm\rangle_{AB}$ have the form, \eqa
 & & |\psi\rangle_C \otimes |\psi^+\rangle_{AB}=
 (1\!\! 1_2 \otimes P - 1\!\! 1_2 \otimes \sigma_3\otimes \sigma_3)
   (|\psi^+\rangle_{CA}\otimes |\psi\rangle_B), \nonumber\\
 & & |\psi\rangle_C \otimes |\psi^-\rangle_{AB}=
 (1\!\! 1_2 \otimes P - 1\!\! 1_8)
   (|\psi^-\rangle_{CA}\otimes |\psi\rangle_B),
\ena in which local unitary transformations of
 $(1\!\! 1_2 \otimes P - 1\!\! 1_2 \otimes \sigma_2\otimes \sigma_2)$
 are exploited,
 \eqa
 & & (1\!\! 1_2 \otimes \sigma_1 \otimes 1\!\! 1_2)
 (1\!\! 1_2\otimes P - 1\!\! 1\otimes \sigma_2\otimes
 \sigma_2) (1\!\! 1_2 \otimes \sigma_1 \otimes 1\!\! 1_2)
 =1\!\! 1_2 \otimes P -1\!\! 1_2 \otimes \sigma_3 \otimes
 \sigma_3,\nonumber\\
 & & (1\!\! 1_2 \otimes i\sigma_2 \otimes 1\!\! 1_2)
 (1\!\! 1_2\otimes P - 1\!\! 1\otimes \sigma_2\otimes
 \sigma_2) (1\!\! 1_2 \otimes i\sigma_2 \otimes 1\!\! 1_2)
 =1\!\! 1_2 \otimes P -1\!\! 1_8.
 \ena

As a remark, a unitary braid is a device of entangling separate
states in the virtual braid teleportation, whereas a unitary
braiding operator plays a role of quantum teleportation in the braid
teleportation.

\subsection{Braid teleportation, crossed measurement and state model}

\begin{figure}
\begin{center}
\epsfxsize=10.5cm \epsffile{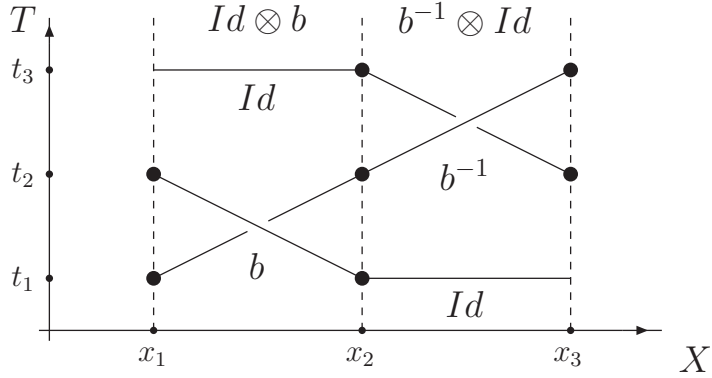} \caption{Braid
teleportation  using the crossed measurement.} \label{fig2}
\end{center}
\end{figure}

 We explore the braid teleportation, ie., a unitary braiding
 operator $(b^{-1} \otimes Id)( Id \otimes b)$, from both physical
 and mathematical perspectives. In view of Vaidman's crossed measurement
 \cite{aav86}, a unitary braid or crossing acts as a device of non-local
 measurement in space-time. In Figure 3, two lines of the crossing $b$
represent two observable operations: the first one relating quantum
measurement at the space-time point $(x_1, t_1)$ to that at the
other point $(x_2, t_2)$ and the second one relating quantum
measurement at $(x_1,t_2)$ to that at $(x_2,t_1)$. The crossed
measurement $(Id\otimes b)$ plays the role of sending a qubit (with
a possible local unitary transformation) from Charlie to Alice.
Similarly, the crossed measurement $(b^{-1}\otimes Id)$ transfers
the qubit from Alice to Bob with a possible local unitary
transformation. Note that this kind of interpretation for a unitary
braiding operator is not the same as the braid statistics of anyons
\cite{wilczek90}.

\begin{figure}
\begin{center}
\epsfxsize=11.cm \epsffile{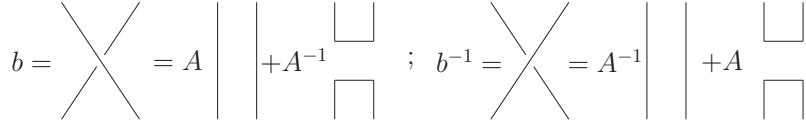} \caption{Braid $b$ and
its inverse $b^{-1}$ in the state model. } \label{fig3}
\end{center}
\end{figure}

\begin{figure}
\begin{center}
\epsfxsize=12.cm \epsffile{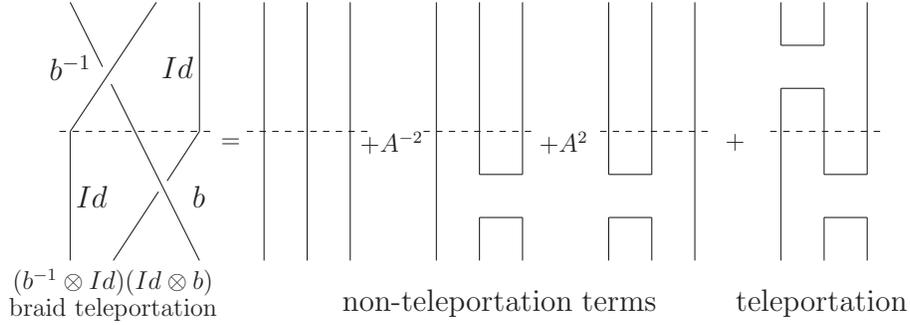} \caption{Braid
teleportation in the state model.} \label{fig4}
\end{center}
\end{figure}

Furthermore, the braid teleportation $(b^{-1} \otimes Id)( Id
\otimes b)$ is different from $( Id \otimes b)(b^{-1} \otimes Id)$,
namely, two crossed measurements are not commutative with each
other. For example, we rewrite the Bell teleportation $(B^{-1}
\otimes 1\!\! 1_2)( 1\!\! 1_2 \otimes B)$ into the other formalism,
 \eqa
 & & (B^{-1} \otimes 1\!\! 1_2)( 1\!\! 1_2 \otimes B)
 =\frac 1 2 ( (1\!\! 1_4 -B^2)\otimes 1\!\! 1_2 ) ( 1\!\! 1_2 \otimes
 (1\!\! 1_4 +B^2)) \nonumber\\
   &=& \frac 1 2 ( 1\!\! 1_2 \otimes (1\!\! 1_4 +B^2) ) (
 (1\!\! 1_4 -B^2)\otimes 1\!\! 1_2) + (1\!\! 1_2\otimes B^2)(B^2\otimes 1\!\! 1_2)
 \nonumber\\
 &=& (1\!\! 1_2\otimes B ) (B^{-1} \otimes 1\!\! 1_2)
 + (1\!\! 1_2\otimes B^2)(B^2\otimes 1\!\! 1_2)
 \ena
where we exploit
 \eq
 (1\!\! 1_2\otimes B^2)(B^2\otimes 1\!\! 1_2)=
 -(B^2\otimes 1\!\! 1_2 ) (1\!\! 1_2\otimes B ).
 \en
and $B^2$ does not form a  braid representation.

 Moreover, we
 explore the configuration for the braid teleportation
$(b^{-1}\otimes Id)(Id \otimes b)$ in the state model for knot
theory \cite{kauffman02}. It is is an approach to a disentanglement
of a knot or link,
 see Figure 4. A braid $b$ is denoted by a under-crossing and its
inverse $b^{-1}$ is denoted by an over-crossing. Each crossing is
identified with a linear combination of two types of configurations:
the first given by two straight lines representing
 identity and the second given by a top cup together with a bottom
cap representing a projector. The coefficients $A, A^{-1}$ are
specified by which state model to be used \cite{kauffman02}. In
Figure 5, the braid teleportation has four diagrammatical terms. A
part above a dashed line is contributed from $(b^{-1}\otimes Id)$
and a part under the dashed line is  from $(Id \otimes b)$.
Obviously, the first three are irrelevant with quantum
 teleportation but the fourth one takes charge for it.

 As a remark, the state model for knot theory \cite{kauffman02} is a
 diagrammatical recipe for the braid representation of the TL
 algebra, and
 the teleportation term in Figure 5 is a typical configuration in
 the diagrammatical TL algebra. In the following, we focus on the
 topic how the TL algebra with local unitary transformations
 describes quantum information and computation.

 \section{Extended TL diagrammatical approach (I):
  diagrammatical rules and examples
          }

In the following sections from Section 3 to Section 7, we propose
the extended Temperley--Lieb diagrammatical approach and exploit it
to study various topics in quantum information and computation. In
this section, we devise the extended TL diagrammatical rules and
explain them by examples.

\subsection{Maximally entangled bipartite pure states}

 Maximally entangled bipartite pure states play key roles in quantum
 information and computation, and how to make a diagrammatical representation
 for them is a bone of the extended TL diagrammatical rules.

The vectors $|e_i\rangle$, $i=0,1,\cdots d-1$ form an orthogonal
basis in a  $d$-dimension Hilbert space $\cal H$ and the covectors
$\langle e_i|$, are chosen in its dual Hilbert space ${\cal
H}^\ast$, \eq
 \sum_{i=0}^{d-1} |e_i\rangle \langle e_i |=1\!\! 1_d, \qquad
 \langle e_j| e_i\rangle =\delta_{ij}, \qquad i,j=0,1,\cdots d-1,
\en where $\delta_{ij}$ is the Kronecker symbol and $1\!\! 1_d $
denotes a $d$-dimensional identity matrix. The maximally entangled
state $|\Omega\rangle$, a quantum state in the two-fold tensor
 product ${\cal H}\otimes {\cal H}$ of the Hilbert space $\cal H$,
 and its dual state $\langle \Omega|$ are respectively given by \eq
 |\Omega\rangle=\frac 1 {\sqrt{d}} \sum_{i=0}^{d-1} |e_i\otimes
 e_i\rangle,
 \qquad \langle \Omega | = \frac 1 {\sqrt{d}} \sum_{i=0}^{d-1}
 \langle e_i\otimes e_i |, \en

The local action of a bounded linear operator $M$ in the Hilbert
space $\cal H$ on $|\Omega\rangle$ satisfies \eqa \label{matrix}
 & &  |\psi\rangle \equiv (M\otimes 1\!\! 1_d) |\Omega\rangle =
 \frac 1 {\sqrt{d}} \sum_{i=0}^{d-1} M|e_i\rangle \otimes |e_i\rangle
 \nonumber\\
 & & = \frac 1 {\sqrt{d}} \sum_{i,j=0}^{d-1} |e_j\rangle M_{ji} \otimes |e_i\rangle
 =\frac 1 {\sqrt{d}} \sum_{i,j=0}^{d-1} |e_j\rangle \otimes
 |e_i\rangle M^T_{ij}
 \nonumber\\
 & & =(1\!\! 1_d \otimes M^T) |\Omega\rangle, \qquad M_{ij}=\langle
 e_i | M |e_j \rangle, \qquad M^T_{ij}=M_{ji},
\ena
 where the upper index $T$ denotes the transpose, and
 hence it is permitted to move a local action of the operator $M$
 in the Hilbert space to the other Hilbert space if it acts on
 $|\Omega\rangle$. A trace of two operators $M^\dag$ and $M^\prime$
 can be represented by an inner product between
 $|\psi\rangle$ and $|\psi^\prime\rangle$,
 \eqa
 \langle \psi| \psi^\prime \rangle &\equiv&\langle \Omega | (M^\dag \otimes 1\!\! 1_d)
 (M^\prime \otimes 1\!\! 1_d) |\Omega\rangle
 = \frac 1 d \sum_{i,j=0}^{d-1} \langle e_i | M^\dag M^\prime
 |e_j \rangle \langle e_i |e_j\rangle, \nonumber\\
 &=& \frac 1 d tr(M^\dag M^\prime), \qquad
 | \psi^\prime \rangle=(M^\prime \otimes 1\!\! 1_d) |\Omega\rangle,
\ena which leads to an inner product with the operator $N_1\otimes
N_2$ given by a trace, \eqa \langle \psi |
 N_1\otimes N_2 |\psi^\prime\rangle =\frac 1 d tr(M^\dag N_1 M^\prime N_2^T).
\ena

The transfer operator $T_{BC}$, sending a quantum state from Charlie
to Bob, is defined by \eq \label{transfer}
 T_{BC} |\psi\rangle_C \equiv   T_{BC} \sum_{k=0}^{d-1} a_k |e_k
 \rangle_C = \sum_{k=0}^{d-1} a_k |e_k\rangle_B = |\psi\rangle_B,
\en and has a form of an inner product between ${}_{CA}\langle
\Omega|$ and $|\Omega\rangle_{AB}$, \eqa
 \label{transfer22}
  {}_{CA}\langle \Omega | \Omega\rangle_{AB} &=& \frac 1 d \sum_{i,j=0}^{d-1}
 ({}_{C}\langle e_i |\otimes {}_A\langle e_i |)
   (|e_j\rangle_A  \otimes |e_j\rangle_B )
 \nonumber\\
 &=& \frac 1 d T_{BC} \equiv \frac 1 d \sum_{i=0}^{d-1}
 |e_i\rangle_B\,\, {}_C\langle e_i |,
\ena which is exploited by Braunstein et al. see \cite{bdms00}.

 A  unitary transformation of $|\Omega\rangle$ is
 called local as the unitary operator $U_n$ only acts on
 the first (or second) Hilbert space of the two-fold Hilbert
 space ${\cal H}\otimes {\cal H}$. The local unitary transformation
 of $|\Omega\rangle$ denoted by
 $|\Omega_n\rangle =(U_n \otimes 1\!\! 1_d)  |\Omega\rangle$, is
 still a maximally entangled vector.  The set of unitary operators
 $U_n$ satisfying the orthogonal relation
  $tr(U_n^\dag U_m)=d\, \delta_{nm}$, $n,m=1\,\cdots d^2$, forms a
  basis of $d\times d$ unitary matrices, where the upper index $\dag$
  denotes the adjoint.
  The collection of maximally entangled states $|\Omega_n\rangle$
  has the  properties, \eq \label{ortho}
  \langle \Omega_n |\Omega_m \rangle =\delta_{nm}, \qquad \sum_{n=1}^{d^2}
  |\Omega_n\rangle \langle \Omega_n|=1\!\! 1_{d^2},
  \qquad n,m=1,\cdots d^2.\en
 Introduce the symbol $\omega_n$ for the maximally entangled state
$|\Omega_n\rangle \langle \Omega_n|$ and especially denote
$|\Omega\rangle\langle \Omega|$ by $\omega$, namely,  \eq
\label{ortho2}
 \omega\equiv|\Omega\rangle \langle \Omega|,
 \qquad \omega_n\equiv|\Omega_n\rangle \langle
 \Omega_n|, \qquad U_1=1\!\! 1_d,
\en and the set of $\omega_n$, $n=1,2,\cdots d^2$ forms a set of
 observables over an output parameter space.

\subsection{Extended TL diagrammatical rules}

\begin{figure}
\begin{center}
\epsfxsize=13.cm \epsffile{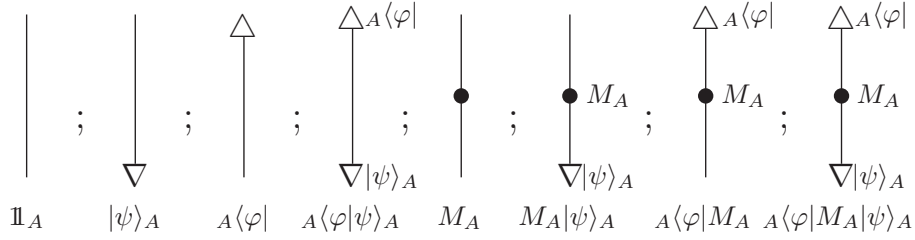} \caption{Straight lines
without or with points.} \label{fig5}
\end{center}
\end{figure}

Three pieces of extended TL diagrammatical rules are devised for
assigning a diagram to a given algebraic object. The first is our
convention; the second explains what straight lines and oblique
lines represent; the third describes various configurations
 in terms of cups and caps.

{\bf Rule 1}. Read an algebraic object such as an inner product from
the left-hand side to the right-hand side, and draw a  diagram from
the top to the bottom. Represent the operator $M$ by a solid point,
its adjoint operator $M^\dag$ by a small circle, its transposed
operator $M^T$ by a solid point with a cross line, and its complex
conjugation operator $M^\ast$ by a small circle with a cross line.
Denote the Dirac ket by the symbol $\nabla$ and the Dirac bra by the
symbol $\triangle$.

\begin{figure}
\begin{center}
\epsfxsize=13.cm \epsffile{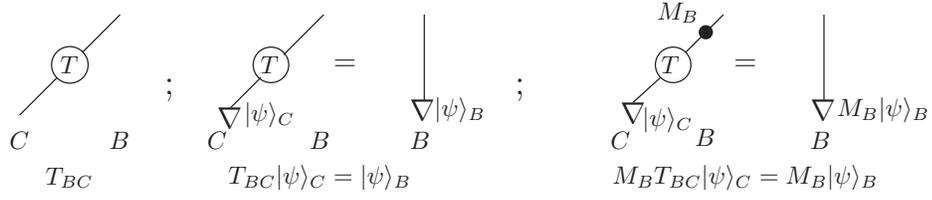} \caption{Oblique lines: the
transfer operator.} \label{fig6}
\end{center}
\end{figure}

{\bf Rule 2}. See Figure 6. A straight line of type $A$ denotes the
identity operator $1\!\! 1_A$ in the system $A$. Straight lines of
type $A$ with a bottom $\nabla$ or top $\triangle$ describe a vector
$|\psi\rangle_A$, covector ${}_A\langle \varphi|$, and an inner
product ${}_A\langle \varphi|\psi\rangle_A$ in the system $A$,
respectively. Straight lines of type $A$ with a middle solid point
or bottom $\nabla$ or
 top $\triangle$  describe an operator $M_A$, a vector $M_A|\psi\rangle_A$,
 a covector ${}_A\langle \varphi| M_A$,  and an inner
 product ${}_A\langle \varphi | M_A |\psi\rangle_A$, respectively.

See Figure 7. An oblique line from the system $C$ to the system $B$
describes the transfer operator $T_{BC}$, and its solid point or
bottom $\nabla$ or top $\triangle$  have the same interpretations as
those on a straight line in Figure 6.

\begin{figure}
\begin{center}
\epsfxsize=13.cm \epsffile{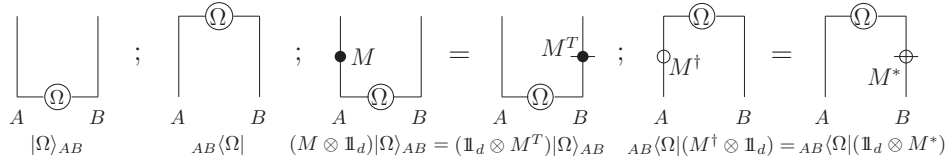} \caption{Cups and caps without
or with points.} \label{fig7}
\end{center}
\end{figure}

{\bf Rule 3}. See Figure 8. A cup denotes the maximally bipartite
entangled state vector $|\Omega\rangle$ and a cap does for its dual
$\langle \Omega|$. A cup with a middle solid point on its one branch
describes a local action of the operator $M$ on $|\Omega\rangle$,
and this solid point can flow to its other branch and then is
replaced by a solid point with a cross line representing $M^T$. The
same happens for a cap except that a solid point is replaced by a
small circle to distinguish the operator $M$ from its adjoint
operator $M^\dag$.

A cup and a cap can form different sorts of configurations. See
Figure 9. As a cup is at the top and a cap is at the bottom  for the
same composite system, this configuration is assigned to the
projector $|\Omega\rangle \langle \Omega |$. As a cap is at the top
and a cup is at the bottom for the same composite system, this
diagram describes an inner product $\langle \Omega|\Omega\rangle=1$
by a closed circle. As a cup is at the bottom for the composite
system ${\cal H}_C \otimes {\cal H}_A$ and a cap is at the top for
the composite system ${\cal H}_A\otimes {\cal H}_B$, that is an
oblique line representing the transfer operator $T_{CB}$ with the
normalization factor $\frac 1 d$.

\begin{figure}
\begin{center}
\epsfxsize=13.5cm \epsffile{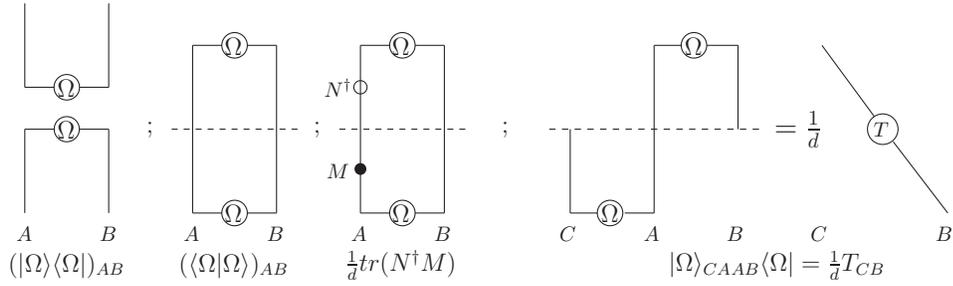} \caption{Three kinds of
combinations of a cup and a cap.} \label{fig8}
\end{center}
\end{figure}

Additionally, as a cup has a local action of the operator $M$ and a
cap has a local action of the operator $N^\dag$, the resulted circle
with a solid point for $M$ and a small circle for $N^\dag$
represents the trace $\frac 1 d tr (M N^\dag)$. As a convention, we
describe a trace of operators by a closed circle with solid points
or small circles, and assign each cap or cup a normalization
 factor $\frac 1 {\sqrt d}$ and a circle a normalization factor $d$.

Note that cups and caps are well known configurations in knot theory
and statistical mechanics. They were used by Wu \cite{wu92} in
statistical mechanics, and exploited by Kauffman \cite{kauffman02}
for diagrammatically representing the Temperley-Lieb algebra soon
after Jones's work \cite{jones87}. These configurations  are
nowadays called Brauer diagrams \cite{brauer37} or Kauffman diagrams
\cite{kauffman02}.

\subsection{Examples for the extended TL diagrammatical rules}

In the following, five examples are listed as well as their
corresponding algebraic counterparts to explain the  extended TL
diagrammatical rules in detail.

Example 1: Figure 10 is a diagrammatical representation for the
teleportation equation (\ref{tele}). The cup denotes the Bell state
$|\phi^+\rangle$, and the cups with middle solid points $\sigma_3$,
$-i\sigma_2$, $\sigma_1$ denote the Bell states $|\phi^-\rangle$,
$|\psi^-\rangle$ and $|\psi^+\rangle$, respectively. The straight
line with a bottom $\nabla$ denotes a unknown state $|\psi\rangle$
to be transported, and other straight lines with middle solid points
respectively denote local unitary transformations of $|\psi\rangle$.
\begin{figure}
\begin{center}
\epsfxsize=12.5cm \epsffile{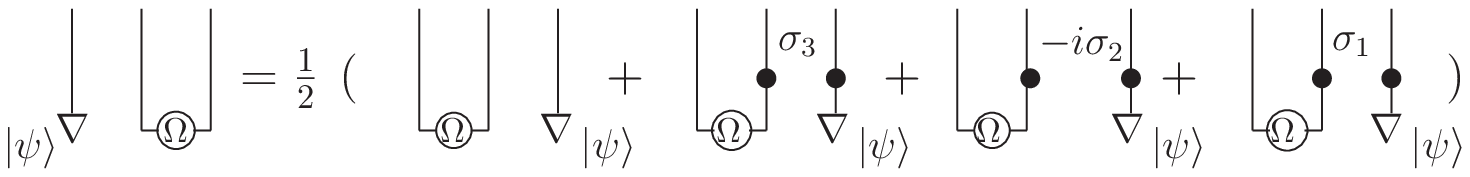} \caption{The
teleportation equation (\ref{tele}).} \label{fig9}
\end{center}
\end{figure}

Example 2: Figure 11 presents how to compute the inner product
between $|\phi\rangle$ and $|\psi\rangle$ in the extended TL
diagrammatical approach. It consists of three terms: the covector
$\langle \phi | \otimes \langle \Omega
 |$, the local operator $1\!\! 1_d  \otimes M \otimes 1\!\! 1_d$, and
 the vector $|\Omega\rangle\otimes |\psi\rangle$. The vector
 $|\psi\rangle$ is represented by a straight line with a bottom $\nabla$,
 and the local operator $1\!\! 1_d \otimes M \otimes 1\!\!
 1_d$ is denoted by a solid point on the cup $|\Omega\rangle$.
 Move the local operator $M$ from its beginning position to the other
 branch of the cap, change it to the local operator $M^T$ and then
 allow the bottom cup and the top cap to collapse into an oblique line
 denoting the transfer operator with a normalization
 factor $\frac 1 d$. Besides, it can be calculated in an algebraic way
 \eqa
 \label{inner}
 & &  \langle \phi \otimes \Omega| (1\!\! 1_d  \otimes M \otimes 1\!\! 1_d)
 |\Omega \otimes \psi \rangle  \nonumber\\
  & & = \frac 1 d \sum_{i,j=0}^{d-1}
  \langle \phi\otimes e_i \otimes e_i |(1\!\! 1_d  \otimes M \otimes 1\!\! 1_d)
 |e_j \otimes e_j \otimes \psi \rangle \nonumber\\
 & & =\frac 1 d \sum_{i,j=0}^{d-1} \langle \phi | e_j \rangle
  \langle e_i | M |e_j \rangle \langle e_i |\psi\rangle
  =\frac 1 d \sum_{i,j=0}^{d-1} \phi^\ast_j M_{ij} \psi_i
  \nonumber\\
 & & = \frac 1 d \sum_{i,j=0}^{d-1} \phi_j^\ast M^T_{ji} \psi_i
 =\frac 1 d \langle \phi | M^T |\psi\rangle
 \ena
in which every step has a diagrammatical counterpart.

\begin{figure}
\begin{center}
\epsfxsize=11.cm \epsffile{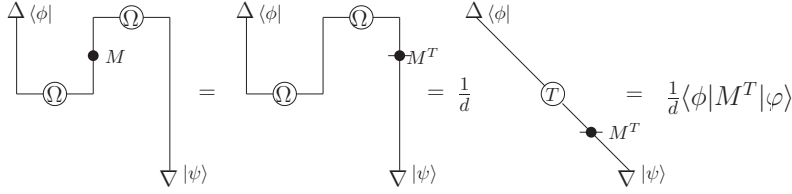} \caption{Inner product in
terms of a cap and a cup.} \label{fig10}
\end{center}
\end{figure}

Example 3:  Figure 12 provides a diagrammatical representation for
the partial trace, which denotes the summation over a subsystem of a
composite system,
 \eq tr_{A}(|e^C_i\otimes e^A_j\rangle\langle e^A_l\otimes e^B_m|)
 =|e^C_i\rangle \langle e^B_m| \delta_{jl},\qquad
 tr_A(|e^A_j\rangle\langle e^A_l |)=\delta_{jl}. \en
Note that the trace is a sort of partial trace, i.e., the summation
over the entire composite system,
 \eq
 tr_{CA} (|e^C_i\otimes e^A_j\rangle\langle e^C_l\otimes
 e^A_m|)=\delta_{il}\delta_{jm}, \qquad i,j,l,m=0,1,\cdots d-1.
 \en
The left diagrammatical term describes the one type of partial trace
leading to a straight line for identity,
 \eq
 tr_A(|\Omega\rangle_{CA}\, {}_{CA}\langle \Omega| )=\frac 1 d
 \sum_{i,j=0}^{d-1} tr_A (|e^C_i\otimes e^A_i\rangle \langle e^C_j \otimes e^A_j|)
 =\frac 1 d (1\!\! 1_d)_C,
 \en
and the other two diagrammatical terms  represent the other type of
partial trace leading to an oblique line for the transfer operator:
the middle term denotes the transfer operator $T_{CB}$ given by
 \eq
 tr_A(|\Omega\rangle_{CA}\, {}_{AB}\langle \Omega| )=\frac 1 d
 \sum_{i,j=0}^{d-1} tr_A (|e^C_i\otimes e^A_i\rangle \langle e^A_j \otimes e^B_j|)
 =\frac 1 d T_{CB},
 \en
 and the right one gives rise to the transfer operator $T_{CB}$ by
 \eq
 tr_A({}_{CA}\langle \Omega |\Omega\rangle_{AB})
 =\frac 1 d T_{BC}.
 \en

\begin{figure}
\begin{center}
\epsfxsize=11.cm \epsffile{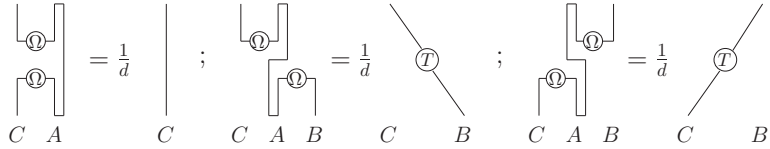} \caption{Three kinds of
partial traces using a cup and a cap.} \label{fig11}
\end{center}
\end{figure}

Example 4: Figure 13 has  two types of diagrams denoting the trace
of operator products. In the first case, a top cup with a bottom cap
forms a same closed circle as a top cap with a bottom cup,
representing an algebraic equation given by \eq
 tr_{CA}( (\rho_C\otimes 1\!\! 1_d |\Omega\rangle_{CA})
 ({}_{CA}\langle \Omega| 1\!\! 1_d\otimes {\cal O}_A^T))
 = {}_{CA}\langle \Omega| (1\!\! 1_d\otimes {\cal O}_A^T)(\rho_C\otimes
1\!\! 1_d) |\Omega\rangle_{CA}, \en
 where $\rho_C$ and ${\cal O}_A$ are bounded linear operators in
 $d$-dimensional Hilbert space.
In the second case, a closed circle formed by two oblique lines for
transfer operators denotes a same trace as a top cap with a bottom
cup, which is algebraically proved,
 \eqa
 \label{oblique}
 tr_{CA} ((\rho_C T_{CA})({\cal O}_A T_{AC})) &=&
 \sum_{i,j=0}^{d-1} tr_{CA} ( (\rho_C|e_i\rangle_C {}_A\langle e_i|)
 ({\cal O}_A|e_j\rangle_A {}_C\langle e_j| ) ) \nonumber\\
 &=&  \sum_{i,j=0}^{d-1} tr_{CA} (
  (\rho_C |e_i\rangle_C\otimes {\cal O}_A |e_j\rangle_A )
  ({}_C\langle e_j|\otimes {}_A\langle e_i|))
  \nonumber\\
  &=& \sum_{i,j=0}^{d-1} (\rho_C)_{ji} ({\cal O}_A)_{ij} =tr(\rho_C {\cal O}_A)
  \nonumber\\
  &=& d\,\cdot {}_{CA}\langle \Omega| (\rho_C\otimes 1\!\! 1_d) (1\!\!
  1_d\otimes {\cal O}^T_A) |\Omega\rangle_{CA}.
 \ena

\begin{figure}
\begin{center}
\epsfxsize=11.cm \epsffile{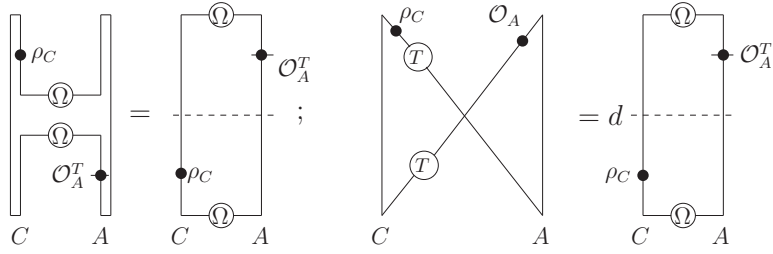} \caption{Closed circles
using cup and cap or oblique lines.} \label{fig12}
\end{center}
\end{figure}

 Example 5:  Figure 14 recognizes the configuration of cup (cap) as
 compositions of cups and caps. In the left diagrammatical term, the
 cup $|\Omega\rangle_{AD}$ is a result of connecting the cap
 ${}_{BC}\langle \Omega|$
 with the cups $|\Omega\rangle_{AB}$ and $|\Omega\rangle_{CD}$,
 which is  verified after some algebra
 \eqa
 & &({}_{BC}\langle \Omega|)(|\Omega\rangle_{AB})(|\Omega\rangle_{CD})
 \nonumber\\
 &\equiv&(1\!\! 1_d\otimes {}_{BC}\langle \Omega|\otimes 1\!\! 1_d)
 (|\Omega\rangle_{AB}\otimes 1\!\! 1_d \otimes 1\!\! 1_d )
 (1\!\! 1_d\otimes 1\!\! 1_d \otimes |\Omega\rangle_{CD})
 \nonumber\\
 &=& \frac 1 {d \sqrt d} \sum_{i,j,k=0}^{d-1} (\langle e^B_i\otimes e^C_i|)
 (|e^A_j\otimes e^B_j \rangle)(|e^C_k\otimes
 e^D_k\rangle)\nonumber\\
  &=& \frac 1 {d \sqrt d}\sum_{i=0}^{d-1} |e^A_i\otimes e^D_i\rangle
  =\frac 1 d |\Omega\rangle_{AD}.
 \ena
The right diagrammatical term shows the cap ${}_{AD}\langle \Omega|$
as a composition of the caps ${}_{AB}\langle \Omega|$,
${}_{CD}\langle \Omega|$ and the cup $|\Omega\rangle_{BC}$,
 \eqa
& ({}_{AB}\langle \Omega|)({}_{CD}\langle
\Omega|)(|\Omega\rangle_{BC}) \nonumber\\
 \equiv&  ({}_{AB}\langle \Omega|\otimes 1\!\! 1_d \otimes 1\!\! 1_d)
 (1\!\! 1_d\otimes 1\!\! 1_d \otimes {}_{CD}\langle\Omega|)
 ( 1\!\! 1_d\otimes |\Omega\rangle_{BC} \otimes 1\!\! 1_d  )
 =\frac 1 d\,\, {}_{AD} \langle \Omega|.
 \ena

\begin{figure}
\begin{center}
\epsfxsize=11.cm \epsffile{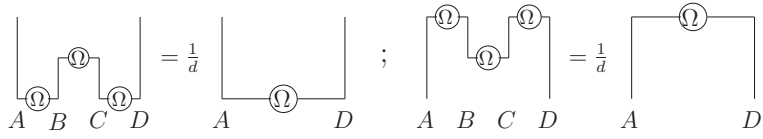} \caption{Cup and cap as
compositions of cups and caps. } \label{fig13}
\end{center}
\end{figure}

As a remark, the above examples will be exploited in the following
sections as topological-like diagrammatical tricks \cite{zk07} in
the extended TL diagrammatical configuration for quantum information
and computation.

 \section{Extended TL diagrammatical approach (II): quantum teleportation}

Three types of descriptions for quantum teleportation: quantum
information flow denoted by the transfer operator, measurement-based
quantum computation, and tight teleportation scheme, are unified in
the extended TL diagrammatical approach.

\subsection{Quantum teleportation using the transfer operator}

 Quantum teleportation can be formulated using the transfer
 operator $T_{BC}$ (\ref{transfer}) which sends the quantum state
 from Charlie to Bob in the way: $T_{BC}
 |\psi\rangle_C=|\psi\rangle_B$, besides its standard description
 \cite{bbcjpw93, bdms00} using the teleportation equation
 (\ref{tele}). The entire teleportation process involves local unitary
 transformations which are not shown in the formalism (\ref{transfer22})
 of the transfer operator $T_{BC}$. To be general, therefore, we recall
 the calculation \cite{preskill} to reformulate the transfer operator
 in terms of maximally entangled states $|\Phi(U) \rangle_{CA}$ and
 $|\Phi(V^T) \rangle_{AB}$ labeled by local unitary actions of $U$ and
 $V^T$ on $|\Omega\rangle$,
 \eqa
 & & {}_{CA} \langle \Phi(U)| \Phi(V^T) \rangle_{AB} \equiv
 {}_{CA} \langle \Omega | ( U^\dag \otimes 1\!\! 1_d  ) ( V^T \otimes 1\!\! 1_d )
 |\Omega \rangle_{AB} \nonumber\\
 & & \equiv  {}_{CAB} \langle \Omega\otimes 1\!\! 1_d |
  ( U^\dag \otimes 1\!\! 1_d \otimes 1\!\! 1_d )
   (1\!\! 1_d \otimes V^T \otimes 1\!\! 1_d )
 |1\!\! 1_d \otimes \Omega \rangle_{CAB} \nonumber\\
 & &= {}_{CAB} \langle \Omega\otimes 1\!\! 1_d |
   (1\!\! 1_d \otimes 1\!\! 1_d \otimes VU^\dag  )
 |1\!\! 1_d \otimes \Omega \rangle_{CAB} \nonumber\\
 & &=\frac 1 d \sum_{i=0}^{d-1} {}_{CB} \langle e_i\otimes 1\!\! 1_d |
   (1\!\! 1_d  \otimes VU^\dag  )
 |1\!\! 1_d \otimes e_i \rangle_{CB} \nonumber\\
 & & \equiv \frac 1 d \sum_{i=0}^{d-1}  {}_C\langle e_i |(VU^\dag)_B
 |e_i\rangle_B = \frac 1 d \sum_{i=0}^{d-1} (VU^\dag)_B
 |e_i\rangle_B\,\, {}_C \langle e_i | \nonumber\\
 & & = \frac 1 d (VU^\dag)_B \,\, T_{BC}
 \ena
which has a special case of $U=V$ given by \eq
 \frac 1 d T_{BC} = {}_{CA} \langle \Phi(U)| \Phi(U^T)\rangle_{AB}.
\en In Figure 15, we repeat the above algebraic calculation  at the
diagrammatical level. From the left to the right, the inner product
${}_{CA} \langle \Phi(U)| \Phi(V^T) \rangle_{AB}$ has the  $\langle
\Omega|$, identity $1\!\! 1_d$, local unitary operators $U$ and
$V^T$, identity $1\!\! 1_d$
 and $|\Omega\rangle$ which are respectively drawn from the
top to the bottom. Move local operators $U^\dag$ and $V^T$ along the
line from their positions to the line denoting the system $B$, and
obtain the product $(VU^\dag)_B$ of local unitary operators acting
on the quantum state that Bob has. The transfer operator $T_{BC}$
has a normalization factor $\frac 1 d$ contributed by vanishing of a
cup and a cap.

\begin{figure}
\begin{center}
\epsfxsize=12.cm \epsffile{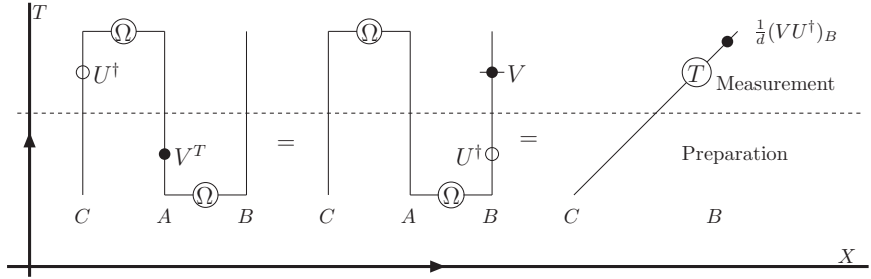} \caption{Quantum
information flow: the transfer operator.} \label{fig14}
\end{center}
\end{figure}

Hence in the extended TL diagrammatical approach, the quantum
teleportation can be viewed as a kind of quantum information flow
denoted by an oblique line from Charlie to Bob. The result $\frac 1
d (VU^\dag)_B T_{BC}$ seems to argue that quantum measurement
labeled by the unitary operator $U^\dag$ plays a role before state
preparation labeled by the unitary operator $V^T$. But it is not
true. Let us read Figure 15 in the way where the $T$-axis denotes
the time-arrow and the $X$-axis denotes the space-distance. Although
the quantum information flow seems to start from the state
preparation, pass through the quantum measurement, and come back to
the state preparation again, and eventually return to the quantum
measurement, it flows from the state preparation to the quantum
measurement without violating the causality principle in its final
form denoted by an oblique line.

 Note that we have to add a rule on how to move operators in the
 extended TL diagrammatical approach: It is forbidden for
 an operator to cross over another operator. For example, we have the
 operator product $\frac 1 d (VU^\dag)_B $ instead of $\frac 1 d
 (U^\dag V)_B$. Obviously, a violation of this rule leads to a violation
 of the causality principle. In addition, there are another
 known approaches to the quantum information flow: the teleportation
 topology \cite{kl04a, kauffman05} and categorical approach
 \cite{coecke04}, which will be compared with the extended
 TL diagrammatical approach in Section 7.

\subsection{Measurement-based quantum teleportation}

Quantum teleportation can be described from the point of quantum
measurement \cite{aav86, vaidman94, vaidman03, eriz05}. The
difference from its standard description \cite{bbcjpw93, bdms00} is
that the maximally entangled state $|\Omega\rangle_{AB}$ between
Alice and Bob is created via quantum measurement \cite{aav86}. Here
it is convenient to represent a quantum measurement in terms of a
projector $(|\Omega\rangle \langle \Omega|)_{AB}$. Therefore,
quantum teleportation is determined by two quantum measurements: the
one denoted by $(|\Omega\rangle \langle \Omega|)_{AB}$ and the other
denoted by $(|\Omega_n\rangle \langle \Omega_n|)_{CA}$, which leads
to a new formulation of the teleportation equation,
 \eq
 \label{mtele}
 (|\Omega_n\rangle \langle \Omega_n|
 \otimes 1\!\! 1_d) (|\psi\rangle \otimes |\Omega \rangle \langle \Omega| )
 =\frac 1 d (|\Omega_n\rangle \otimes 1\!\! 1_d )
  (1\!\! 1_d \otimes (1\!\! 1_d \otimes U_n^\dag
  |\psi\rangle) \langle \Omega| ),  \en where
  lower indices $A, B, C$ are omitted for convenience. Read this teleportation
  equation (\ref{mtele}) from the left to the right,
  and draw the diagram from the top to the bottom in view of the extended TL
  diagrammatical rules, i.e., Figure 16.

\begin{figure}
\begin{center}
\epsfxsize=11.0cm \epsffile{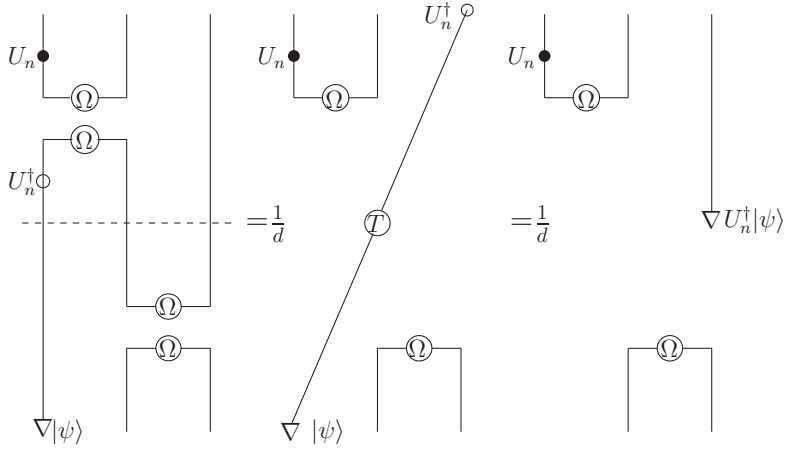}
\caption{Measurement-based quantum teleportation.} \label{fig15}
\end{center}
\end{figure}

There is a natural connection between two formalisms (\ref{tele})
and (\ref{mtele}) of the teleportation equation. Consider all
unitary matrices $U_n$ satisfying (\ref{ortho}) and (\ref{ortho2}),
and then make a summation of all teleportation equations of the type
(\ref{mtele}) labeled by $U_n$ to derive
 \eq
 |\psi \rangle \otimes |\Omega\rangle = \frac 1 d \sum_{n=1}^{d^2}
  |\Omega_n\rangle \otimes U_n^\dag |\psi\rangle
 \en
 which is a generalization of the teleportation equation of the
 type (\ref{tele}) in the $d$-dimension Hilbert space.
 As $d=2$, the collection of the basis of unitary operators
  consist of unit matrix $1\!\! 1_2$ and Pauli matrices $\sigma_1, \sigma_2,
  \sigma_3$. Bell measurements are denoted by projectors
  in terms of Bell states $|\phi^\pm\rangle$ and $|\psi^\pm\rangle$.
  They lead to the same kind of the teleportation equations
  of the type (\ref{mtele}),
 \eqa
 & & (|\phi^- \rangle \langle \phi^- | \otimes 1\!\! 1_2  )
   (|\psi\rangle \otimes |\phi^+\rangle   )
   = \frac 1 2 ( |\phi^- \rangle  \otimes  \sigma_3 |\psi\rangle ),
    \nonumber\\
 & &  (|\psi^+ \rangle \langle \psi^+ | \otimes 1\!\! 1_2  )
   (|\psi\rangle \otimes |\phi^+\rangle  )
   = \frac 1 2 ( |\psi^+ \rangle  \otimes \sigma_1 |\psi\rangle),
    \nonumber\\
 & &  (|\psi^- \rangle \langle \psi^- | \otimes 1\!\! 1_2  )
   (|\psi\rangle \otimes |\phi^+\rangle )
   = \frac 1 2 ( |\psi^- \rangle  \otimes  -i \sigma_2 |\psi\rangle), \nonumber\\
& & (|\phi^+ \rangle \langle \phi^+ | \otimes 1\!\! 1_2  )
   (|\psi\rangle \otimes |\phi^+\rangle  ) =
    \frac 1 2 (|\phi^+ \rangle  \otimes |\psi\rangle),
   \ena
 which has a summation as the teleportation equation (\ref{tele})
 since Bell states $|\phi^\pm\rangle$ and $|\psi^\pm\rangle$
 satisfy
  \eq
 1\!\! 1_{2} =|\phi^+\rangle \langle \phi^+ |  + |\phi^-\rangle \langle
 \phi^-| +|\psi^+\rangle \langle \psi^+|  + |\psi^-\rangle \langle
 \psi^-|.
  \en

 Furthermore, we comment on measurement-based quantum teleportation using
 continuous variables \cite{vaidman94}, which is a simple generalization
 of the discrete teleportation without essential conceptual changes.
 A maximally entangled state $|\Omega^\prime\rangle$ and teleportated
 state $|\Psi\rangle$ in the continuous case have the form,
 \eq
 |\Omega^\prime\rangle =\int d x\,\, |x,x\rangle, \qquad |\Psi\rangle=\int d
 x\,\, \psi(x)\,\, |x\rangle,
 \en
 and other maximally entangled states
 $|\Omega_{\alpha\beta}^\prime\rangle$ are formulated by the combined
 action of the $U(1)$ rotation with the translation $T$
 on $|\Omega^\prime\rangle$,
 \eq
 |\Omega_{\alpha\beta}^\prime\rangle = (U_\beta \otimes T_\alpha ) |\Omega^\prime\rangle
 \equiv \int d x \exp(i\beta x) |x,x+\alpha \rangle, \qquad \alpha,
 \beta \in {\mathbb R}
 \en
  where $U_\beta|x\rangle=e^{i\beta x}|x\rangle$,
  $T_\alpha|x\rangle=|x+\alpha\rangle$  and which is a common eigenvector of
  the location operator
 $\mathbf{X}\otimes 1\!\! 1- 1\!\! 1\otimes \mathbf{X}$ and
 conjugate momentum operator
 $\mathbf{P} \otimes 1\!\! 1 +1\!\! 1 \otimes \mathbf{P}$,
 \eq
 (\mathbf{X}\otimes 1\!\! 1- 1\!\! 1\otimes \mathbf{X})|\Omega^\prime_{\alpha\beta}\rangle=
 -\alpha |\Omega^\prime_{\alpha\beta}\rangle, \qquad
 (\mathbf{P} \otimes 1\!\! 1 +1\!\! 1 \otimes \mathbf{P})|\Omega^\prime_{\alpha\beta}\rangle
 = 2 \beta|\Omega^\prime_{\alpha\beta}\rangle.
 \en
The teleportation equation of the type (\ref{mtele}) is obtained to be \eq
  (|\Omega^\prime_{\alpha\beta}\rangle  \langle\Omega^\prime_{\alpha\beta}| \otimes 1\!\!
  1 ) (|\Psi \rangle \otimes |\Omega^\prime\rangle  )
  =(|\Omega^\prime_{\alpha\beta}\rangle  \otimes 1\!\! 1 )
  (1\!\! 1 \otimes 1\!\! 1 \otimes   U_{-\beta} T_{\alpha}
  |\Psi\rangle  )
  \en
 which has a similar diagrammatical representation as Figure 16. The
 translation operator $T_\alpha$ is its own adjoint operator, and hence it
 is permitted to move along a cup or cap without change.

 As a remark, the difference between Figure 15 and 16 lies in there
 are an extra cup and cap besides the quantum information flow in
 Figure 15, which become crucial when we study the tight
 teleportation scheme in the next subsection. In addition,
 Figure 16 is a typical configuration in the
 diagrammatical representation for the TL algebra as
  involved local unitary operators are identity, see Section 5.

\subsection{Tight teleportation and dense coding schemes}

\begin{figure}
\begin{center}
\epsfxsize=13.5cm \epsffile{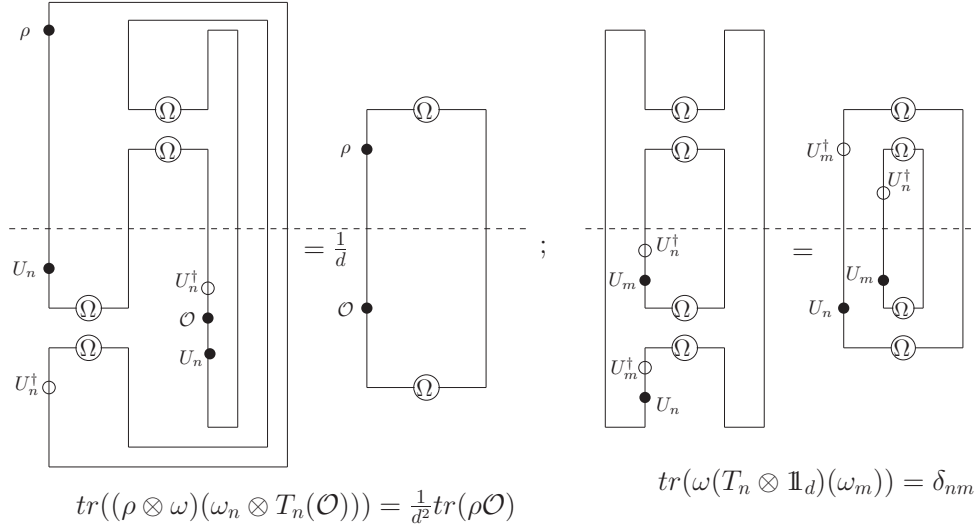} \caption{Characteristic
equations for tight teleportation and dense coding.  } \label{fig16}
\end{center}
\end{figure}

 In the tight teleportation and dense coding schemes \cite{werner01},
 all involved finite Hilbert spaces are  $d$ dimensional and
the classical channel distinguishes $d^2$ signals. All examples we
treated in the above belong to the tight class. We exploit the same
notations as in \cite{werner01}. The density operator $\rho$ is a
positive operator with a normalized trace. Charlie has his density
operator $\rho_C=(|\phi_1\rangle \langle \phi_2 |)_C$ which denotes
the quantum state to be sent to Bob. Alice and Bob share the
maximally entangled state $\omega_{AB}$. The set of observables
$\omega_n$, $n=1,2,\cdots d^2$ over an output parameter space  is a
collection of bounded linear operators in the Hilbert space $\cal
H$. Alice performs Bell measurements in the composite system between
Charlie and her, and she chooses her observables $(\omega_n)_{CA}$
as local unitary transformations of the maximally entangled state
$\omega$. After Bob gets the message denoted by $n$ from Alice, he
applies a local unitary transformation $T_n$ on his observable
${\cal O}_B$, given by
 \eq
 T_n ({\cal O}_B ) =U_n^\dag {\cal O}_B U_n, \qquad {\cal O}_B =(|\psi_1
 \rangle \langle \psi_2 |)_B, \qquad n=1,2, \cdots d^2.
\en
 The operator $T_n$ defined this way is called a channel, a complete
 positive linear operator with the normalization $T_n(1\!\! 1_d)=1\!\! 1_d$.

In terms of $\rho_C$, $\omega_{AB}$, $(\omega_n)_{CA}$ and
$T_n({\cal O}_B)$, the tight teleportation scheme is summarized in
the equation
 \eq
 \label{ttele}
 \sum_{n=1}^{d^2} tr ( (\rho\otimes \omega ) (\omega_n \otimes T_n({\cal O}) ))
 =tr(\rho {\cal O}),
 \en
where lower indices $A, B, C$ are neglected and which is called the
characteristic equation for quantum teleportation in our previous
work \cite{zk07}. It catches the aim of a successful teleportation:
Charlie performs quantum measurement in his system as he does in
Bob's system, though they are far away from each other.  The term
containing the message $n$ has a form denoted by $term_n$ in the
following \eqa
 & & {term}_n =tr ( ( |\phi_1\rangle \langle \phi_2| \otimes
  |\Omega\rangle \langle \Omega |) ( |\Omega_n\rangle \langle \Omega_n | \otimes
   U_n^\dag | \psi_1\rangle \langle \psi_2| U_n ) )
   \nonumber\\
   & &=\langle \Omega_n \otimes \psi_2 U_n |\phi_1 \otimes
   \Omega\rangle \langle \phi_2 \otimes \Omega|\Omega_n \otimes
   U_n^\dag \psi_1 \rangle \nonumber\\
   & &= (\frac 1 d \langle \psi_2 | \phi_1\rangle)
     (\frac 1 d \langle \phi_2 | \psi_1\rangle)
 = \frac 1 {d^2} tr (\rho {\cal O}),
\ena where the inner product (\ref{inner}) is applied twice. There
are $d^2$ distinguished messages labeled by $n$, so we prove the
characteristic equation (\ref{ttele}). In the left term of Figure
17, we have two ways of deriving the characteristic equation
(\ref{ttele}) at the diagrammatical level. The first moves local
operators $U_n^\dag$, ${\cal O}$ and $U_n$ from the branch of the
cap to the other branch and then applies known diagrammatical tricks
by the first term of Figure 12 and the first term of Figure 13. The
second makes use of tricks by the second and third terms of Figure
12 and the second term of Figure 13. In addition, the number of
classical channel, $d^2$ counts all possible teleportation diagrams
of the same type.

In view of tight dense coding schemes have the same elements as the
tight teleportation,  all the tight dense coding schemes
\cite{werner01} are concluded in the characteristic equation, \eq
\label{dense}
 tr(\omega(T_n\otimes 1\!\! 1_d)(\omega_m))=\delta_{nm}
\en which is explained as follows. As Alice and Bob share the
maximally entangled state $|\Omega\rangle_{AB}$, she transforms her
state by the channel $T_n$ to encode the message $n$, and then Bob
performs  the measurement on observables $\omega_m$ in his system.
At $n=m$, Bob gets the message. The entire process of dense coding
is performed in the way
 \eqa
 & & tr( |\Omega\rangle \langle \Omega | (U_n^\dag \otimes 1\!\! 1_d )
 |\Omega_m\rangle \langle \Omega_m | (U_n \otimes 1\!\! 1_d) )\
 \nonumber\\
 & &=\langle \Omega| U_n^\dag \otimes 1\!\! 1_d |\Omega_m\rangle
 \langle \Omega_m | U_n \otimes 1\!\! 1_d |\Omega\rangle
 =\frac 1 {d^2} ( tr(U_n^\dag U_m))^2 =\delta_{nm}
 \ena
which derives the characteristic equation for tight dense coding
(\ref{dense}), also proved in the right term of Figure 17.

As a remark, Figure 16 denoting measurement-based quantum
computation includes all elements of quantum teleportation: Figure
15 representing the quantum information flow is its part, and the
left term of Figure 17 is its closure. Therefore, three approaches
to quantum teleportation are unified in the extended TL
diagrammatical approach.

 \section{Extended TL diagrammatical approach (III):
   TL algebra and Brauer algebra}

We study algebraic structures in the extended TL diagrammatical
approach: the TL algebra, the Brauer algebra, and the extended TL
category. Note that symbols $\Omega$ labeling maximally entangled
states  are omitted for convenience without confusion in Figures 18,
19, 21.

 \subsection{Diagrammatical representation of TL algebra}

 The TL algebra $TL_n(\lambda)$ over the complex field $\mathbb{C}$
 is generated by identity $Id$ and $n-1$
 Hermitian projectors $e_i$ satisfying \eqa \label{tl}
 e_i^2 &=&  e_i, \qquad (e_i)^\dag=e_i,\,\,\, i=1,\ldots,n-1, \nonumber\\
 e_i e_{i\pm1} e_i &=& \lambda^{-2} e_i, \qquad e_i e_j=e_j e_i, \,\,\, |i-j|>1,
  \ena
with $\lambda$ called loop parameter \footnote{The notation $e_i$
for an idempotent of the TL algebra is easily confused with a basis
vector $|e_i\rangle$. As they both appear in the same subsection, we
denote $|e_i\rangle$ by $|i\rangle$.}.  The diagrammatical
 representation of the TL algebra  is called
 the Brauer diagram \cite{brauer37} or Kauffman diagram \cite{kauffman02}
 in literature. It is a planar $(n,n)$ diagram including a ``hidden" rectangle
 in the plane with $2 n$ ``hidden" distinct points: $n$ points on its top edge
 and $n$ points on its bottom edge, and they are connected by disjoint strings
 drawn in the rectangle. The identity is a diagram with all strings vertical,
 and $e_i$ has its $i$th and $i+1$th top (bottom)
 boundary points connected with all other strings vertical.
 The multiplication $e_i e_j$ identifies bottom points of $e_i$
with corresponding top points of $e_j$, removes a common boundary,
and replaces each obtained loop with a factor $\lambda$. The adjoint
of $e_i$ is its image under a mirror reflection on a horizontal
line.

 \begin{figure}
 \begin{center}
 \epsfxsize=13.5cm \epsffile{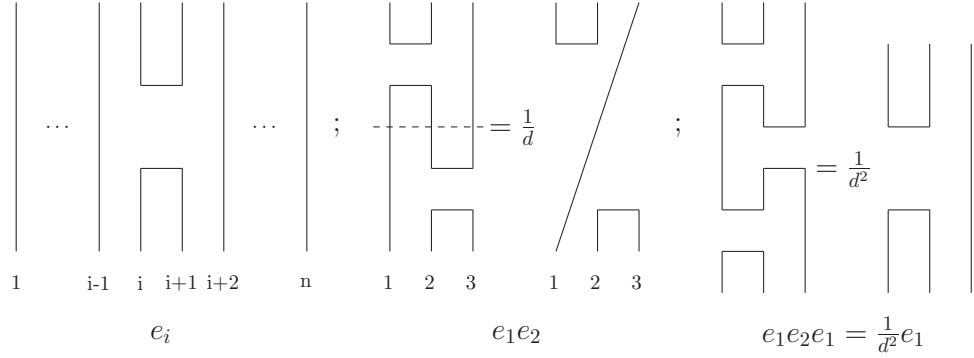} \caption{Generator $e_i$,
 multiplication $e_1 e_2$ and the axiom for the $TL_3(d)$ algebra.}
 \label{fig17}
 \end{center}
 \end{figure}

Let us set up a representation of the $TL_n(d)$ algebra in terms of
the maximally entangled state $\omega$,  \eq \omega=|\Omega\rangle
\langle \Omega |=\frac 1 d \sum_{i,j=0}^{d-1}|ii\rangle \langle j
j|, \quad \omega^2=\omega \en by defining idempotents $e_i$ as \eq
e_i=(Id)^{\otimes (i-1)} \otimes \omega \otimes (Id)^{\otimes
(n-i-1)}, \quad i=1,\cdots n-1 \en with loop parameter $d$. For
example, a representation of the $TL_3(d)$ algebra is generated by
two idempotents $e_1$ and $e_2$, \eq
 e_1=\omega\otimes Id, \qquad e_2=Id\otimes \omega,
\en and they proved to satisfy the axiom $e_1 e_2 e_1 =\frac 1 {d^2}
e_1$ in the way \eq
 e_1 e_2 e_1 |\alpha\beta\gamma\rangle =\frac 1 d \sum_{l=0}^{d-1}
 e_1 e_2 |ll\gamma\rangle \delta_{\alpha\beta}=\frac 1 {d^3}
 \sum_{n=0}^{d-1} |nn\gamma\rangle \delta_{\alpha\beta}=\frac 1
 {d^2} e_1 |\alpha\beta\gamma\rangle
 \en
 as well as the axiom $e_2 e_1 e_2 = \frac 1 {d^2} e_2$ using
 similar calculation. Figure 18 is a diagrammatical representation
 for $e_i$, $e_1 e_2$ and $e_1 e_2 e_1 =\frac 1 {d^2} e_1$ with
 loop parameter $d$. Therefore, a cup (cap) introduced in
 the extended TL diagrammatical approach is a connected line
 between top (bottom) boundary points. Each cup (cap) with a
 normalization factor $d^{-\frac 1 2}$ leads to an additional
 normalization factor $d^{-\frac 1 2 N}$ as the number of
vanishing cups and caps is $N$, and a closed circle yields a
normalization factor $d=tr(1\!\! 1_d)$. For examples, $e_1 e_2$ has
a normalization factor $\frac 1 d$ from a vanishing cup and  cap,
and $e_1 e_2 e_1 $ has a factor $\frac 1 {d^2}$ from four vanishing
cups and caps.

\begin{figure}
\begin{center}
\epsfxsize=13.5cm \epsffile{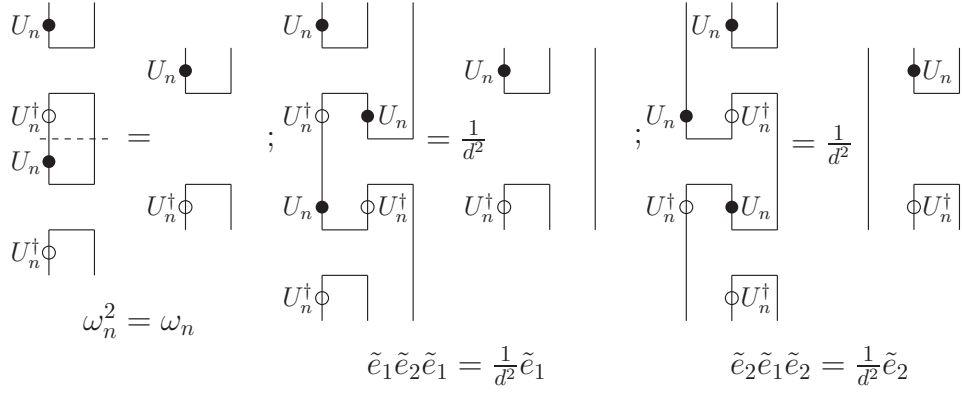} \caption{The $TL_3(d)$
algebra in the extended TL diagrammatical approach.} \label{fig18}
\end{center}
\end{figure}

In terms of the maximally entangled state $\omega_n$ as a local
unitary transformation $U_n$ of $\omega$, we can set up a
representation of the $TL_n(d)$ algebra, too. For example, a
representation of the $TL_3(d)$ algebra is generated by
 $\tilde{e}_1$ and $\tilde{e}_2$,
 \eq \tilde{e}_1=\omega_n\otimes Id, \qquad \tilde{e}_2=Id \otimes
\omega_n, \en which are proved to satisfy the axioms of the TL
algebra in the extended TL diagrammatical approach, see Figure 19.
Hence the axioms of the TL algebra are invariant under local unitary
transformations as idempotents $e_i$ are generated by the maximally
entangled state $\omega$.

\begin{figure}
\begin{center}
\epsfxsize=13.5cm \epsffile{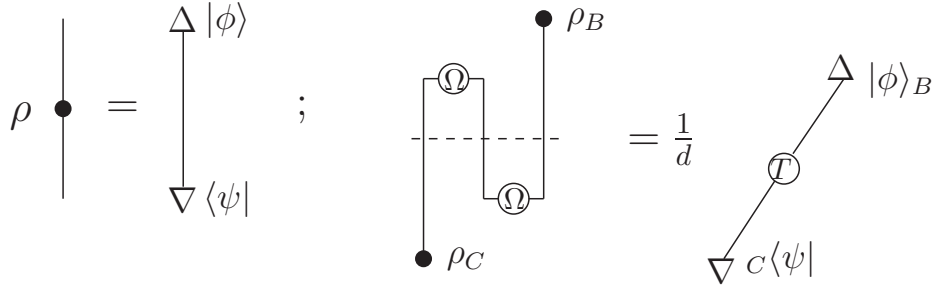} \caption{Quantum
information flow with the density operator.} \label{fig19}
\end{center}
\end{figure}

Furthermore, a representation of the $TL_3(d)$ algebra is
constructed in the way \eq
 e^\prime_1 = \rho\otimes \omega, \qquad e^\prime_2=\omega\otimes \rho, \qquad
 \rho=|\phi\rangle \langle \psi|, \,\, tr(\rho)=1.
\en because $\rho$ and $\omega$  and the tensor products between
them are all projectors. The axioms of the TL algebra are checked in
both algebraic and diagrammatical approaches. Do calculation
 \eqa
& & (\omega\otimes \rho)(\rho \otimes \omega)
 =\frac 1 {d^2} \sum_{i,j=0}^{d-1} (|ii\rangle\langle j j|\otimes
  |\phi\rangle \langle \psi|)
  \sum_{l,m=0}^{d-1} ( |\phi\rangle \langle \psi|\otimes |ll\rangle \langle m m | )
  \nonumber\\
  & &= \frac 1 {d^2} \sum_{i,j=0}^{d-1}  \sum_{l,m=0}^{d-1}
  (|ii \phi\rangle\langle j j \psi|)
 ( |\phi l l\rangle \langle \psi m m|)
  =  \frac 1 {d^2} \sum_{i,j=0}^{d-1} |ii \phi\rangle\langle \psi j j| \ena
which leads to a proof for the axiom $e_2^\prime e_1^\prime
e_2^\prime=\frac 1 {d^2} e_2^\prime$, \eqa
 & & (\omega\otimes \rho) (\rho \otimes \omega) (\omega\otimes \rho)
 =\frac 1 {d^3}  \sum_{i,j=0}^{d-1}  \sum_{l,m=0}^{d-1}
  |ii \phi\rangle \langle \psi j j|l l \phi\rangle
  \langle m m \psi| \nonumber\\
 & & = \frac 1 {d^2} \sum_{i,j=0}^{d-1} |ii \phi\rangle\langle  j j \psi|
    =\frac 1 {d^2} (\omega\otimes \rho).
\ena Similarly to prove $e^\prime_1 e^\prime_2 e^\prime_1=\frac 1
{d^2} e^\prime_1$. As a remark, $\rho=|\phi\rangle \langle \psi|$ so
that the transfer operator $T_{BC}$ sends a half of $\rho_C$,
$|\phi\rangle_C$ from Charlie to Bob to form a unit inner product
with ${}_B\langle \psi|$, a half of $\rho_B$, see Figure 20.
Moreover, in terms of $\rho$ and $\omega_n$, a representation of the
$TL_3(d)$ algebra is also set up in another way \eq
\tilde{e}^\prime_1=\rho\otimes \omega_n, \qquad
\tilde{e}^\prime_2=\omega_n\otimes \rho, \qquad n=1,\cdots d^2,\en
if and only if a local unitary transformation $U_n$ is a symmetric
matrix:
 \eq
 U_n^T = U_n, \qquad U_n^T U_n^\dag= U_n^\ast U_n=1\!\! 1_d
\en which can be observed in Figure 19.

 \subsection{Quantum Teleportation: the Brauer algebra}

 Quantum teleportation has a same diagrammatical representation as the
 element $e_1 e_2$ ($e_2 e_1$) of the TL algebra, if involved local
 unitary transformations are identity (see Figure 16). There exists a
 natural question which can be possibly asked. Quantum teleportation
 plays important roles in quantum information, but the product
 $e_1 e_2$ ($e_2 e_1$) is only an element of the TL algebra.
 It is meaningful to explore in which case the configuration
 of $e_1 e_2$ ($e_2 e_1$) is crucial in the mathematical sense.
 In the following, we present the axioms of the Brauer algebra
 \cite{brauer37} and explain that the quantum teleportation
 configuration forms a bone of this algebra.

 \begin{figure}
\begin{center}
\epsfxsize=13.5cm \epsffile{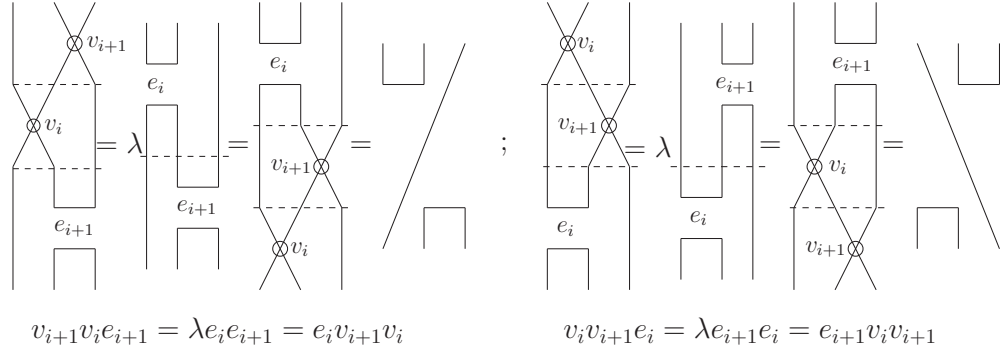} \caption{Quantum
teleportation: the Brauer algebra.} \label{fig25}
\end{center}
\end{figure}

 The Brauer algebra $D_n(\lambda)$
 is an extension of the TL algebra with
 virtual crossings, $\lambda$ called loop parameter. It has two
 types of generators: idempotents $e_i$ of the TL algebra $TL_n(\lambda)$
 satisfying (\ref{tl}) and virtual crossings $v_i$ satisfying (\ref{vbgr1}),
 $i=1,\cdots n-1$. Both generators  satisfy the mixed relations of
 the Brauer algebra,
 \eqa
 \label{brauer}
 (ev/ve): \,\,\, e_i v_i =v_i e_i =e_i, \qquad e_i v_j =v_j e_i, \qquad j\neq i\pm 1,
 \nonumber\\
 (vve):\,\,\, v_{i\pm1}v_i e_{i\pm1}=\lambda e_i e_{i\pm1}, \qquad
 (evv):\,\,\, e_i v_{i\pm1} v_i =\lambda e_i e_{i\pm1}
 \ena which has a diagrammatical representation in Figure 21.
 For example,  the permutation $P$ as a virtual crossing and the
 maximally entangled state $\omega$ as an idempotent
\eq
 P=\sum_{i,j=0}^{d-1} |i\otimes j\rangle \langle j\otimes i |,\qquad
 \omega=\frac 1 d \sum_{i=0}^{d-1} |i\otimes i\rangle \langle j \otimes j |
\en
 form a representation of the Brauer algebra $D_2(d)$
 with loop parameter $d$.
  The axiom $(ev/ve)$ is verified in the way \eq P\omega=\frac 1 d \sum_{i,j=0}^{d-1}
\sum_{i^\prime,j^\prime=0}^{d-1}
 |i\otimes j\rangle \langle j\otimes i |i^\prime\otimes
 i^\prime\rangle \langle j^\prime \otimes j^\prime |=\omega=\omega P,
\en and the axioms $(vve)$ and $(eev)$ are proved after some algebra
\eqa
 & &(1\!\! 1_d \otimes P)(P\otimes 1\!\! 1_d)(1\!\! 1_d \otimes \omega)
 = d (\omega \otimes 1\!\! 1_d)(1\!\! 1_d \otimes \omega)=
 (\omega\otimes 1\!\! 1_d)(1\!\! 1_d\otimes P)(P\otimes 1\! \! 1_d), \nonumber\\
 & &(P\otimes 1\!\! 1_d ) (1\!\! 1_d\otimes P) (\omega\otimes 1\!\! 1_d )
 = d (1\!\! 1_d \otimes \omega ) (\omega \otimes 1\!\! 1_d ) = (1\!\! 1_d \otimes \omega )
  (P\otimes 1\!\! 1_d ) (1\!\! 1_d\otimes P ), \nonumber
\ena which are also proved in Figure 21 with $\lambda=d$.

As a remark on Figure 21, it is clear that the configuration for
quantum teleportation is fundamental for defining the Brauer
algebra.  The Brauer algebra presents an equivalent approach of
performing the teleportation using the swap gate $P$ and Bell
measurement, where the teleportation swapping $(P\otimes
Id)(Id\otimes P)$ or $(Id\otimes P)(P\otimes Id)$ are involved. As a
summary of algebraic structures underlying quantum teleportation, we
have proposed the braid teleportation, the teleportation swapping,
the virtual braid teleportation, the Temperley--Lieb algebra, and
the Brauer algebra.

 \subsection{Comment on the extended TL category}

 In this paper, we propose the extended TL diagrammatical approach
 to quantum information and computation involving maximally
 entangled states and local unitary transformations. Its
 diagrammatical configurations consist of cups, caps, solid points,
 empty circles, etc., as an extension of Kauffman diagrams or Brauer
 diagrams. An extension of the TL algebra with local unitary
 transformations is called {\em the extended TL algebra}, and the
 collection of all extended TL algebras is called {\em the extended TL
 category} which contains abundant mathematical objects
 such as  braids, permutation, the TL algebra, the Brauer algebra,
 and others (see the next section). This category is a mathematical
 foundation of our diagrammatical approach.

 Furthermore, interested readers are
 invited to refer our previous work \cite{zk07}, in which {\em unitary
 Hermitian ribbon categories} are suggested as a mathematical
 description of quantum information and physics as well as the extended TL
 diagrammatical approach is viewed as a diagrammatical
 representation for tensor categories.

 \section{Extended TL diagrammatical approach (IV):
 entanglement swapping and universal quantum computation}

 We study the entanglement swapping,
 universal quantum computing, and multipartite entanglements,
 in the extended TL diagrammatical approach which is a
 diagrammatical representation for the extended TL category.

  \subsection{Entanglement swapping}

   Entanglement swapping \cite{ekert93} is an experimental technique
   realizing quantum entanglement between two independent systems as a
   consequence of quantum measurement instead of physical
   interaction. Let us make an example for its theoretical
   interpretation in terms of a projector representing quantum measurement.
   Alice has a bipartite entangled state $|\Omega_l\rangle^{A}_{ab}$
   for particles $a, b$, and Bob has $|\Omega_m\rangle^B_{cd}$ for particles
   $c,d$. They are independently created and do not share
   common history. Alice applies quantum measurement denoted by
   $Id\otimes |\Omega_n\rangle \langle \Omega_n|\otimes Id$
   to the product state of $|\Omega_l\rangle^{A}_{ab}$ and
   $|\Omega_m\rangle^B_{cd}$, so that the output called the
   entanglement swapped state $|\Omega_{lnm}\rangle^{AB}_{ad}$ is
   a bipartite entangled state shared by Alice and Bob for particles $a,d$,
    \eqa
   \label{swap}
  & & (Id\otimes |\Omega_n\rangle \langle \Omega_n|\otimes Id)
  (|\Omega_l\rangle^{A}_{ab} \otimes |\Omega_m\rangle^B_{cd})
  \nonumber\\
  & &= \frac 1 d ( Id\otimes |\Omega_n\rangle \otimes Id)
  \frac 1 {\sqrt d}\sum_{i=0}^{d-1}
  ( U_l U_n^\ast U_m  |e_i\rangle^A_a\otimes Id
  \otimes Id\otimes |e_i\rangle^B_d)  \nonumber\\
  & &\equiv \frac 1 d ( Id\otimes |\Omega_n\rangle \otimes Id)\,\,
   |\Omega_{lnm}\rangle^{AB}_{ad}.
    \ena
In other words, the entanglement swapping reduces a four-particle
state $|\Omega_l\rangle^{A}_{ab}\otimes |\Omega_m\rangle^B_{cd}$ to
a bipartite entangled state $|\Omega_{lnm}\rangle^{AB}_{ad}$ using
entangling measurement. As a remark, the entanglement swapped state
$|\Omega_{lnm}\rangle^{AB}_{ad}$ plays a  role in the comparison of
quantum mechanics with classical physics, because it is a quantum
state but produced without any classical physical interactions.

\begin{figure}
\begin{center}
\epsfxsize=12.5cm \epsffile{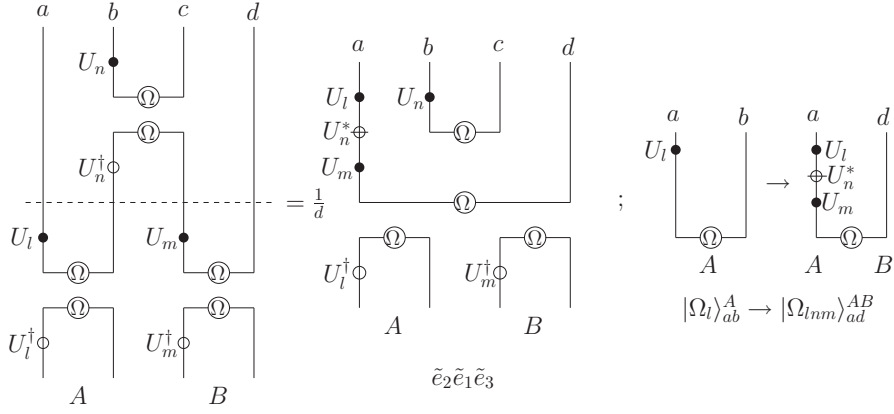}
 \caption{Entanglement swapping in the extended TL category.}
 \label{fig20}
\end{center}
\end{figure}

Read the entanglement swapping equation (\ref{swap}) from the left
to the right and draw a diagram from the top to the bottom according
to the extended TL diagrammatical rules, i.e., the left term of
Figure 22. It is a configuration for an element $\tilde{e}_2
\tilde{e}_1 \tilde{e}_3$ in the extended TL category, i.e., \eq
 \tilde{e}_2 \tilde{e}_1 \tilde{e}_3=(Id\otimes \omega_n \otimes Id)
 (\omega_l \otimes Id \otimes Id)(Id\otimes Id \otimes
 \omega_m).
\en which changes the entangled state $|\Omega_l\rangle^{A}_{ab}$ in
Alice's system to the entangled state
 $|\Omega_{lnm}\rangle^{AB}_{ad}$  in Alice
and Bob's composite system, see the right term of Figure 22.
Furthermore, the entanglement swapping is also called the
teleportation using a cup state.
  Alice measures the Bell state $|\Omega\rangle_{AB}$
  with a projector
  $|\psi\rangle_A \langle \psi |$ so that she transfers her
  quantum state $|\psi\rangle_A$ to Bob in the way
 \eqa
  && |\psi\rangle_A\langle \psi |\Omega\rangle_{AB}
  = \frac 1 {\sqrt d} |\psi\rangle_A
  \sum_{i,j=0}^{d-1} {}_{AB}\langle e_j\otimes Id | \psi^\ast_j |e_i
  \otimes e_i \rangle_{AB} \nonumber\\
 & &  = \frac 1 {\sqrt
 d} |\psi\rangle_A \sum_{i=0}^{d-1}  \psi^\ast_i |e_i
 \rangle_{B} =\frac 1 {\sqrt d} |\psi\rangle_A (\langle \psi |)_B^T,
  \ena
 which has a diagrammatical representation, the left term of Figure
 23 with $(\langle e_i|)^T=|e_i\rangle$, similar to the two-way
 teleportation in the crossed measurement \cite{vaidman94, eriz05}.

\begin{figure}
\begin{center}
\epsfxsize=12.cm \epsffile{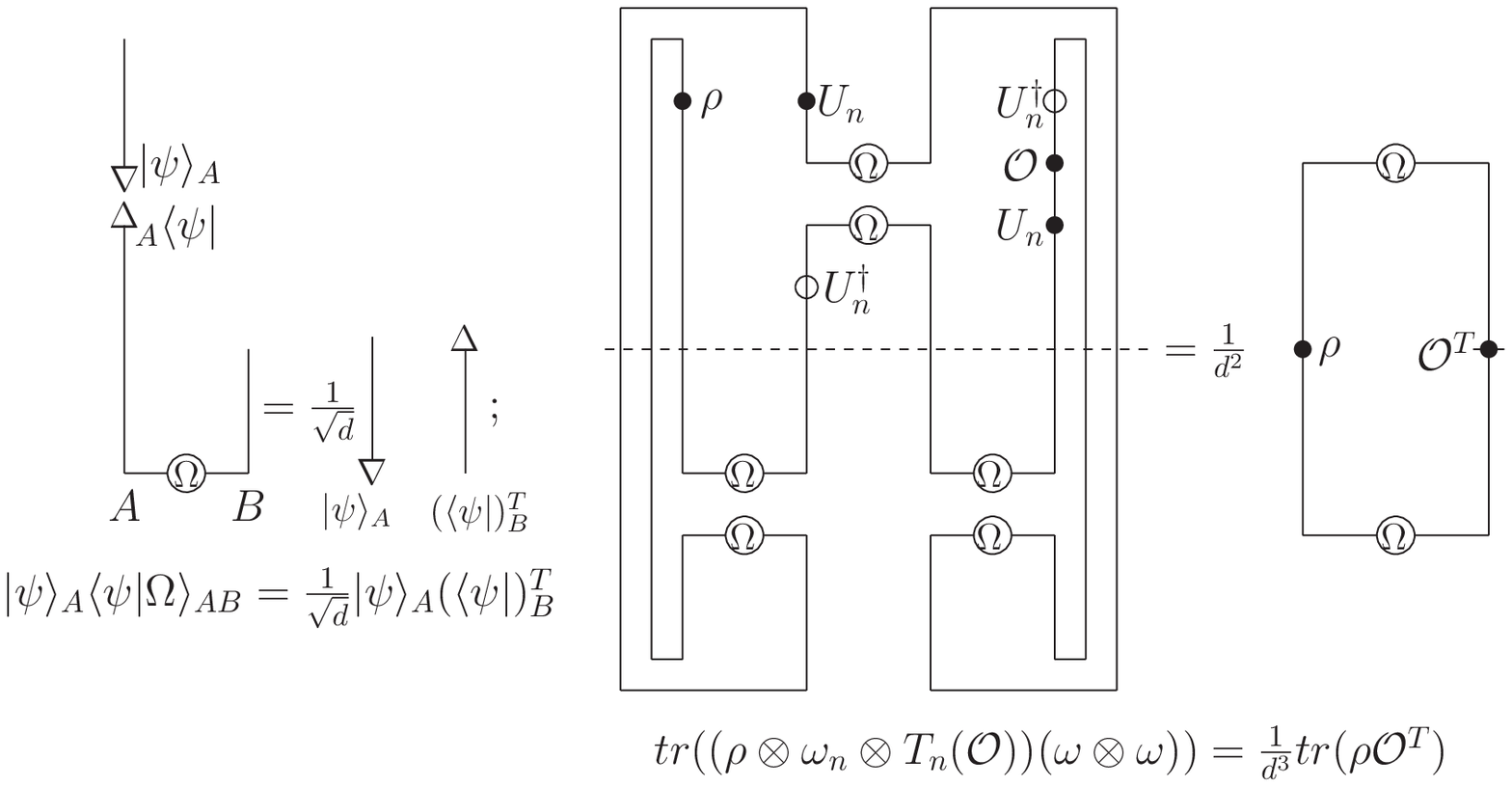} \caption{Tight
entanglement swapping in the extended TL category.} \label{fig21}
\end{center}
\end{figure}

The characteristic equation for the tight entanglement swapping
 is derived by a similar procedure of obtaining characteristic equations
 for tight teleportation and dense coding schemes \cite{werner01},
\eq \label{tightswap}
 \sum_{n=1}^{d^2} tr ((\rho\otimes \omega_n \otimes T_n({\cal O}))
 (\omega\otimes \omega))
 =\frac 1 d tr (\rho {\cal O}^T)
\en where the density operator $\rho$ for particle $a$, observable
${\cal O}$ for particle $d$, and quantum channel $T_n({\cal O})$ for
particle $d$ are respectively given by \eq \
 \rho=|\phi_1\rangle \langle \phi_2|, \,\,\,
 {\cal O}=|\psi_1\rangle \langle \psi_2|, \,\,\,
 T_n({\cal O})=U_n^\dag {\cal O} U_n \en and the transpose
 of the density operator, ${\cal O}^T$ is defined in the way
 \eq {\cal O}^T=\sum_{i,j=0}^{d-1} \psi_{1i}
 \psi_{2j}^\ast (|e_i\rangle \langle e_j|)^T=\sum_{i,j=0}^{d-1}
 \psi_{1i} \psi_{2j}^\ast |e_j\rangle \langle e_i|,\,\, (\langle
 e_i|)^T=|e_i\rangle. \en
 The tight entanglement swapping equation (\ref{tightswap}) is
 easily proved in the extended TL diagrammatical approach, see
 the right term of Figure 23 which is a closure of the left term
 of Figure 22. The diagrammatical trick by
 Figure 14 is exploited to derive the same configuration
 as the first term of Figure 13.
 It can be also verified in an algebraic way: the $term_n$
 given by
 \eqa
 & & term_n\equiv tr ((|\phi_1\rangle \langle \phi_2 |\otimes |\Omega_n\rangle
 \langle \Omega_n|\otimes U_n^\dag|\psi_1\rangle \langle \psi_2| U_n )
  (|\Omega\rangle \langle \Omega| \otimes |\Omega\rangle \langle \Omega|) )
 \nonumber\\
 & &= \langle \phi_2 \otimes \Omega_n \otimes \psi_2
 U_n|\Omega\otimes \Omega\rangle \langle \Omega\otimes
 \Omega|\phi_1\otimes \Omega_n\otimes U_n^\dag \psi_1 \rangle
 \nonumber\\
 & & = \frac 1 {d^3} (\phi_2^\ast\cdot \psi^\ast_2) (\phi_1\cdot\psi_1)
 \ena
is found to be
 \eq
 \frac 1 {d^3} tr(\rho {\cal O}^T)
 =\frac 1 {d^3} \sum_{i,j=0}^{d-1} \psi_{1i} \psi^\ast_{2j}
 \langle e_i |\phi_1\rangle \langle \phi_2 |e_j\rangle
=term_n, \en and the characteristic equation (\ref{tightswap}) id
proved due to there are $d^2$ $term_n$ with each $term_n$
independent of $n$.

Therefore, the extended TL diagrammatical approach is not only to
describe a quantum information protocol but also to assign it a
characteristic equation.

 \begin{figure}
\begin{center}
\epsfxsize=12.cm \epsffile{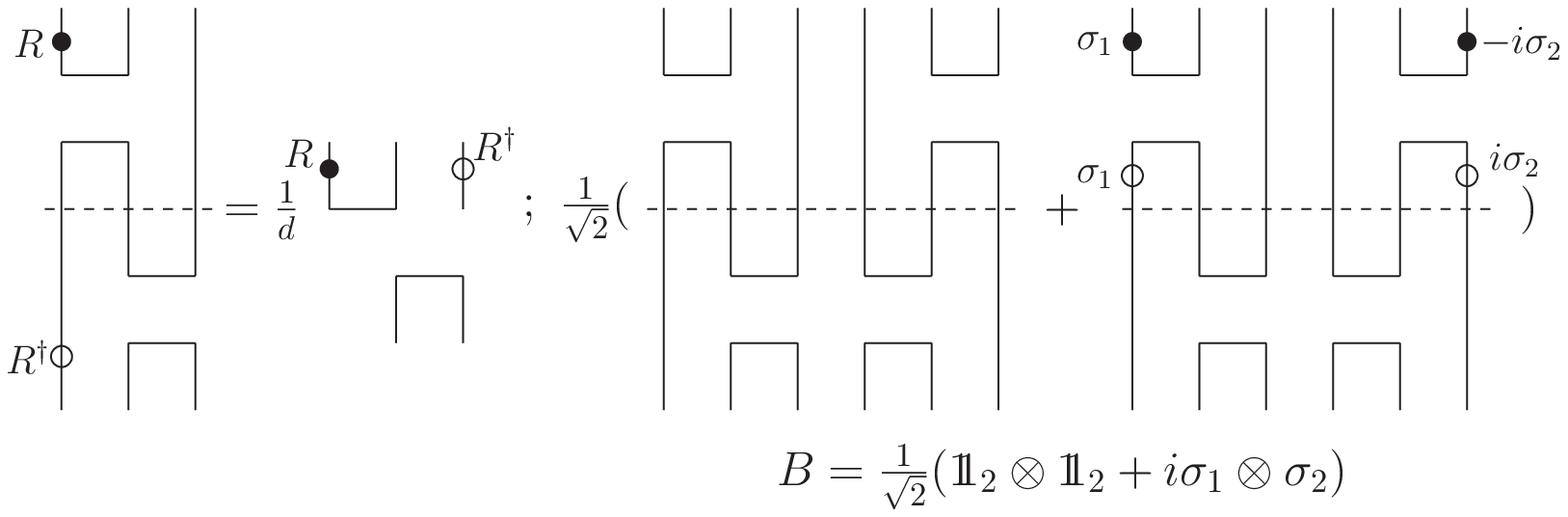} \caption{Single-qubit
gate and unitary braid gate.} \label{fig22}
\end{center}
\end{figure}

 \subsection{Universal quantum computing}

 Quantum teleportation has been considered as a universal quantum
 computational primitive \cite{gc99}. Under such a proposal, there
 are both theoretical observations and experimental motivations.
 The teleported state permits the action of local unitary
 transformations, and so quantum teleportation realizes single-qubit
 gates as local unitary transformations and two-qubit gates as linear
 combinations of products of single-qubit gates. Besides single-qubit
 transformations and Bell measurements can be performed in labs.
 Additionally, we have fault-tolerant quantum computation
 \cite{shor96, preskill97}, as single-qubit transformations are performed
 fault-tolerantly.

 \begin{figure}
 \begin{center}
 \epsfxsize=12.5cm \epsffile{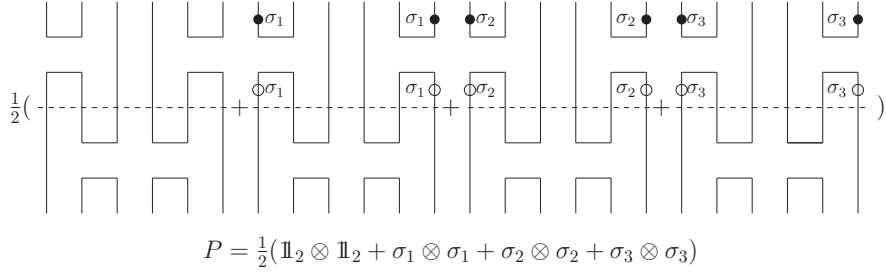} \caption{Swap gate in the
 extended TL category} \label{fig23}
 \end{center}
 \end{figure}

 A fault-tolerant gate $R$, an element of the
 Clifford  group \cite{shor96, preskill97}, enters the teleportation
 via entangling measurement and then is transported in the form of its
 conjugation $R^\dag$, see the left term of Figure 24. A unitary
 braid gate (\ref{bell}) has a form given by
 \eq
B=\frac 1 {\sqrt 2} (1\!\! 1_2 \otimes 1\!\! 1_2 + i \sigma_1\otimes
\sigma_2)
 \en
 which is performed  in the way shown in the extended TL diagrammatical
 approach, see the right term of Figure 24.
 It has two diagrammatical terms and each one consists of
 two teleportation processes for sending a two-qubit.
 The swap gate $P$, denoted by (\ref{perm}),
 is an element of the extended TL category, i.e.,
Figure 25 which has four diagrammatical terms and each one
represents an algebraic term in (\ref{perm}). The CNOT gate, a
linear combination of products of Pauli matrices,
 \eqa
 CNOT &=&(|0\rangle \langle 0|\otimes 1\!\! 1_2 +|1\rangle \langle 1|\otimes \sigma_1 )
 = \frac 1 2 (1\!\! 1_2+\sigma_3 ) \otimes 1\!\! 1_2 + \frac 1 2 (1\!\! 1_2 -\sigma_3)
 \otimes \sigma_1, \nonumber\\
 &=& \frac 1 2 (1\!\! 1_2 \otimes 1\!\! 1_2 +1\!\! 1_2\otimes \sigma_1
 +\sigma_3\otimes 1\!\! 1_2  -\sigma_3\otimes \sigma_1) \ena satisfying basic
properties of the CNOT gate, $$
 CNOT|00\rangle=|00\rangle, \,\, CNOT|01\rangle=|01\rangle,\,\,
 CNOT|10\rangle=|11\rangle,\,\, CNOT|11\rangle=|10\rangle,  $$
 is an element of the extended TL category,
 Figure 26. Note that symbols $\Omega$ labeling a cup and cap
 are omitted in Figures 24-26 for convenience.

 \begin{figure}
\begin{center}
\epsfxsize=12.5cm \epsffile{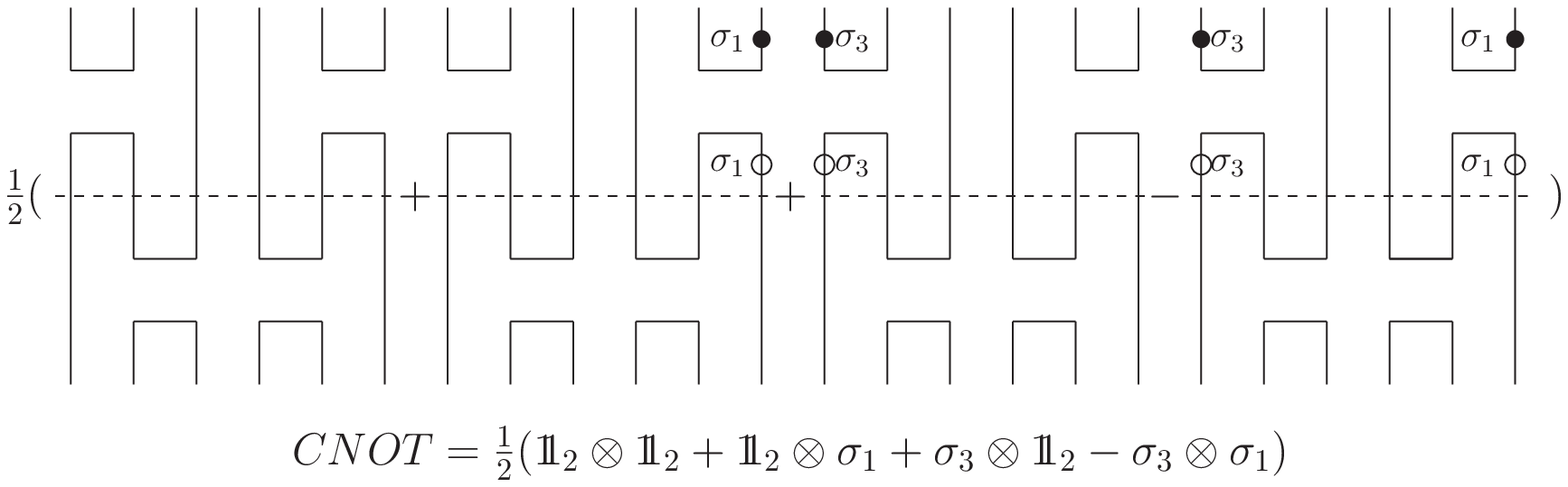} \caption{CNOT gate in the
extended TL category.} \label{fig24}
\end{center}
\end{figure}

Knot polynomial in terms of a unitary braid gate \cite{kl04a,
zkg05a, zkg05b} in the extended TL category can be computed using
quantum simulation of knot on quantum computer, which is different
from an approximate quantum algorithm  \cite{aharonov05} for
computing the Jones polynomial as well as topological quantum
computing \cite{kitaev97,flw02} involving unitary braid
representations as quantum gates acting on quasi-particles like
anyons. Furthermore, virtual knots can be simulated via a quantum
program with unitary braid gates and swap gates. Moreover, exactly
solvable two dimensional quantum field theories or statistical
models \cite{yang67, baxter72} can be simulated on quantum computer,
since unitary solutions of the Yang--Baxter equation with spectral
parameters \cite{zkg05a,zkg05b} can be performed in the extended TL
category.

As a remark, the extended TL category is a low-dimensional
``topological" model for universal quantum computation, and
``topological" or topological-like features \cite{zk07} are expected
to be helpful to look for new quantum algorithms.

  \subsection{Comment on multipartite entanglements}

 Bell
 measurements and local unitary transformations are crucial elements for the
 application of the extended TL diagrammatical rules.
 Hence, multipartite maximally entangled states like the GHZ state or the
 state $|\chi\rangle$ can be treated in the extended TL category if they have a form
 in terms of Bell measurements and local unitary transformations.
 For example, the GHZ state $|GHZ\rangle$ is a linear combination of local unitary
 transformations of Bell state,
 \eqa
 |GHZ\rangle & =& \frac 1 {\sqrt 2} (|0\rangle\otimes |00\rangle
     + |1\rangle\otimes |11\rangle)
 \nonumber\\
 &=& \frac 1 2 (|0\rangle + |1\rangle) \otimes |\phi^+\rangle +\frac 1 2
  (|0\rangle-|1\rangle)\otimes |\phi^-\rangle \nonumber\\
  &=& \frac 1 2 (1\!\! 1_8 + \sigma_3\otimes 1\!\! 1_2 \otimes \sigma_3 )
  (|\alpha\rangle\otimes |\phi^+\rangle)
 \ena
 where $|\alpha\rangle=|0\rangle + |1\rangle$. Similarly,
 the four-particle state $|\chi\rangle$ \cite{gc99},  in the construction
  of the CNOT gate using quantum teleportation, has a form \eqa
 & |\chi\rangle =\frac 1 {\sqrt 2} (|00\rangle+|11\rangle) |00\rangle +\frac 1 {\sqrt 2}
 (|01\rangle+|10\rangle) |11\rangle \nonumber\\
 &= \frac 1 {\sqrt 2} |\phi^+\rangle (|\phi^+\rangle+|\phi^-\rangle )
  +\frac 1 {\sqrt 2}
 |\psi^+\rangle (|\phi^+\rangle -|\phi^-\rangle) \nonumber\\
 &= \frac 1 {\sqrt 2} (1\!\! 1_{16} + 1\!\! 1_8\otimes \sigma_3
 +1\!\! 1_2\otimes \sigma_1\otimes 1\!\! 1_4
   -1\!\! 1_2\otimes \sigma_1 \otimes 1\!\! 1_2\otimes \sigma_3 )
 |\phi^+\rangle |\phi^+\rangle.
\ena

 We will explore quantum teleportation using multipartite
 maximally entanglement states in the extended TL category in our
 further research. As a remark, the extended TL diagrammatical approach
 rules can be applied to topics like Bell inequalities, quantum
 cryptography and so on, in which Bell measurements and local unitary
 transformations play fundamental roles.

 \section{Extended TL diagrammatical approach (V):
  quantum information flow}

\begin{figure}
\begin{center}
\epsfxsize=10.cm \epsffile{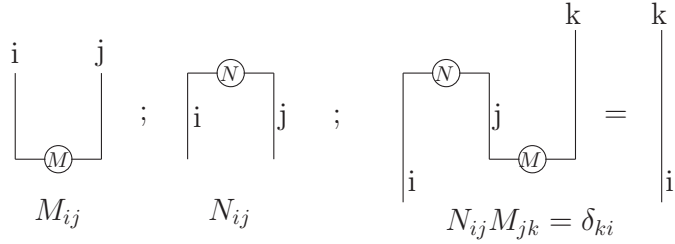} \caption{Teleportation
topology: the topological condition.} \label{fig26}
\end{center}
\end{figure}

 We describe the quantum information flow in the extended TL
 category, and compare this description with another two known
 approaches: the teleportation topology \cite{kl04a, kauffman05}
 and strongly compact closed categories \cite{coecke04}.

 \subsection{Teleportation topology}

 Teleportation topology \cite{kl04a,kauffman05} explains quantum
 teleportation as a kind of topological amplitude satisfying the
 topological condition. There are one-to-one correspondences
 between quantum amplitude and topological amplitude. The state
 preparation describes a creation of a two-particle quantum state
 from vacuum with a diagrammatical representation denoted by a cup state
 $|Cup\rangle$, and quantum measurement denotes an annihilation of a
 two-particle quantum state with a diagrammatical representation
 by a cap state  $\langle Cap|$. See Figure 27. The cup and cap states
 are associated with the matrices $M$ and $N$ in the way \eq
 |Cup\rangle = \sum_{i,j=0}^{d-1} M_{ij} |e_i\otimes e_j\rangle,
 \qquad \langle Cap | = \sum_{i,j=0}^{d-1} \langle e_i\otimes e_j |
 N_{ij}
 \en which satisfy the topological condition, i.e., the
 concatenation of a cup and a cap is a straight line denoted by the
 identity matrix $N_{ij} M_{jk}=\delta_{ik}$.

 In the extended TL diagrammatical approach, the concatenation of
 a cup and a cap is formulated by the transfer operator which is not
 identity required by the topological condition, see Figure 15.
 Besides, the cup and
 cap states are normalized maximally entangled states given by
 \eq
 |Cup\rangle =\frac 1 {\sqrt{d}} \sum_{i=0}^{d-1}  |e_i\otimes e_i\rangle,
 \qquad \langle Cap | = \frac 1 {\sqrt{d}} \sum_{i=0}^{d-1}
   \langle e_i\otimes e_i |
 \en which assigns  a normalization factor $\frac 1 d$ to a straight
 line from the concatenation of a cup and a cap. Furthermore, a
 projector formed by a top cup and bottom cap forms a representation
 of the TL algebra. Moreover, quantum teleportation is the quantum
 information flow denoted by
 the topological condition in the teleportation topology, whereas it is
 described in Figure 16 with the quantum information flow as its part.

 \subsection{Quantum information flow in terms of maps}

\begin{figure}
\begin{center}
\epsfxsize=12.5cm \epsffile{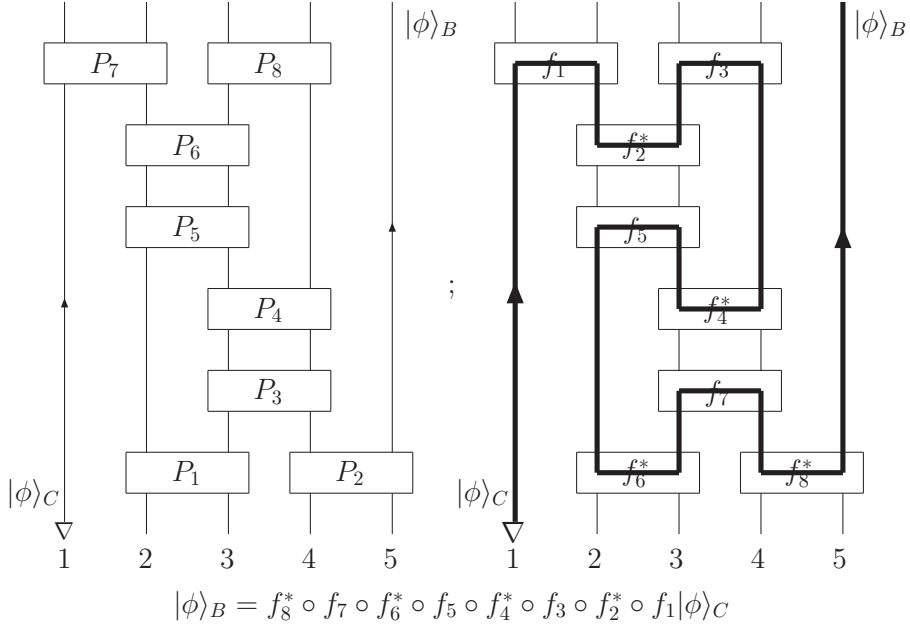} \caption{Quantum
information flow in the categorical approach.} \label{fig27}
\end{center}
\end{figure}

Quantum teleportation is an information protocol transporting a
unknown quantum state from Charlie to Bob with the help of Alice. To
describe it in a unified mathematical formalism, we have to
integrate standard quantum mechanics with classical features, since
the outcomes of measurements are sent to Bob from Alice via
classical channels and then Bob performs a required unitary
operation. The categorical approach proposed by Abramsky and Coecke
describes the quantum information flow by strongly compact closed
categories, see \cite{coecke04, ac04, coecke05}.

To sketch the quantum information flow in the form of a composition
of a series of maps which are central topics of the category theory,
we study an example in detail. Set five Hilbert spaces ${\cal H}_i$
and its dual ${\cal H}^\ast_i$, $i=1,\cdots,5$, and define eight
bipartite projectors $P_\alpha=| \Phi_\alpha \rangle \langle
\Phi_\alpha |$, $\alpha=1, \cdots, 8$ in which the bipartite vector
$|\Phi_\alpha\rangle$ is an element of ${\cal H}_i \otimes {\cal
H}_{i+1}$, $i=1,\cdots, 4$. In the left diagram of Figure 28, every
box represents a bipartite projector $P_\alpha$, and the vector
$|\phi\rangle_C\in {\cal H}_1$ that Charlie owns is transported to
Bob who obtains the vector $|\phi\rangle_B \in {\cal H}_5$ through
the quantum information flow. The projectors $P_1$ and $P_2$ pick up
an incoming vector in ${\cal H}_2 \otimes {\cal H}_3 \otimes {\cal
H}_4 \otimes {\cal H}_5 $, and the projectors $P_7$ and $P_8$
determine an outgoing vector in $ {\cal H}_1 \otimes  {\cal H}_2
\otimes {\cal H}_3 \otimes {\cal H}_4 $. The right diagram in Figure
28 shows the quantum information flow from $|\phi\rangle_C$ to
$|\phi\rangle_B$. It is drawn according to permitted and forbidden
rules \cite{coecke04}: the flow is forbidden to go through a box
from the one side to the other side, and is forbidden to be
reflected at the incoming point, and has to change its direction
from an incoming flow to an outgoing flow as it passes through a
box. Obviously, if these rules are not imposed there will be many
possible paths from $|\phi\rangle_C$ to $|\phi\rangle_B$.

Let us set up one-to-one correspondence between a bipartite vector
and a map. There are a $d_1$-dimension Hilbert space ${\cal
H}_{(1)}$ and a $d_2$-dimension Hilbert space ${\cal H}_{(2)}$. The
bipartite vector $|\Phi\rangle$ has a form in terms of the product
basis $|e^{(1)}_i\rangle\otimes |e^{(2)}_j\rangle $ in ${\cal
H}_{(1)} \otimes {\cal H}_{(2)}$,
 \eq
 |\Phi\rangle=\sum_{i=0}^{d_1-1}\sum_{j=0}^{d_2-1} m_{ij} |e^{(1)}_i\rangle\otimes
|e^{(2)}_j\rangle, \qquad \langle \Phi | =
\sum_{i=0}^{d_1-1}\sum_{j=0}^{d_2-1} m^\ast_{ij} \langle e^{(1)}_i
|\otimes \langle e^{(2)}_j |
 \en
 where $\langle \Phi|$ denotes the dual vector of $|\Phi\rangle$ in
 the dual product space ${\cal H}^\ast_{(1)} \otimes {\cal
 H}^\ast_{(2)}$ with the basis $\langle e^{(1)}_i |\otimes \langle e^{(2)}_j |
 $. Besides, once the product basis is fixed,
 bipartite vectors $|\Phi\rangle$ or $\langle \Phi |$
 are  determined by a $d_1\times d_2$ matrix $M_{d_1\times
 d_2}=(m_{ij})$. Defining two types of maps $f$ and $f^\ast$ in the way
 \eqa
 & & f: {\cal H}_1 \rightarrow {\cal H }^\ast_2, \qquad f(\cdot)=
 \sum_{i=0}^{d_1-1}\sum_{j=0}^{d_2-1} m_{ij} \langle
 e^{(1)}_i|\cdot\rangle \langle e^{(2)}_j |, \nonumber\\
 & & f^\ast: {\cal H}^\ast_1 \rightarrow {\cal H }_2, \qquad f^\ast(\cdot)=
 \sum_{i=0}^{d_1-1}\sum_{j=0}^{d_2-1} m_{ij}  |e^{(2)}_j \rangle \langle
 \cdot| e^{(1)}_i \rangle,
 \ena
we have the bijective correspondences,
 \eq
 |\Phi\rangle \approx \langle \Phi | \approx M \approx f \approx
 f^\ast
 \en
which suggests that the bipartite project box in Figure 28 can be
labeled by the map $f$ or $f^\ast$ or matrix $M$.

Now we work out the formalism of the quantum information flow in the
categorical approach. Consider a projector $P_7=|\Phi_7\rangle
\langle \Phi_7 | $ and introduce a map $f_1$ to represent the action
of $\langle \Phi_7 |$, a half of $P_7$,
 \eq
f_1: {\cal H}_1 \rightarrow {\cal H}_2^\ast, \qquad
f_1(\phi_C)=\langle \Phi_7 |\phi\rangle_C.
 \en
Similarly, the remaining seven boxes are respectively labeled by the
maps $f_2^\ast$, $f_3$, $f_4^\ast$, $f_5$, $f_6^\ast$, $f_7$ and
$f_8^\ast$, defined by
 \eqa
 & & f^\ast_2: {\cal H}^\ast_2 \rightarrow {\cal H}_3, \qquad
  f_2^\ast \circ f_1 (\phi_C) = \langle \Phi_7 |\phi_C \otimes
  \Phi_6\rangle, \nonumber\\
 & & f_3: {\cal H}_3 \rightarrow {\cal H}^\ast_4, \qquad
  f_3\circ f_2^\ast \circ f_1 (\phi_C) = \langle \Phi_7\otimes \Phi_8 |\phi_C \otimes
  \Phi_6\rangle, \nonumber\\
  & & f_4^\ast: {\cal H}^\ast_4 \rightarrow {\cal H}_3, \qquad
  f_4^\ast \circ f_3\circ f_2^\ast \circ f_1 (\phi_C)
  = \langle \Phi_7\otimes \Phi_8 |\phi_C \otimes
  \Phi_6 \otimes \Phi_4\rangle, \nonumber\\
  & &  f_5: {\cal H}_3 \rightarrow {\cal H}^\ast_2, \qquad
  f_5\circ f_4^\ast \circ f_3\circ f_2^\ast \circ f_1 (\phi_C) =
   \langle \Phi_5\otimes \Phi_7\otimes \Phi_8 |\phi_C \otimes
  \Phi_6 \otimes \Phi_4\rangle, \nonumber\\
 & &  f_6^\ast: {\cal H}_2^\ast \rightarrow {\cal H}_3, \qquad
  f_7: {\cal H}_3 \rightarrow {\cal H}_4^\ast,
  \qquad  f^\ast_8: {\cal H}_4^\ast \rightarrow {\cal H}_5, \nonumber\\
 & & f_6^\ast \circ f_5\circ f_4^\ast \circ f_3\circ f_2^\ast \circ f_1
 (\phi_C) =\langle \Phi_5 \otimes \Phi_7\otimes \Phi_8 |\phi_C \otimes
  \Phi_6 \otimes \Phi_4 \otimes \Phi_1\rangle, \nonumber\\
 & &
  f_7\circ f_6^\ast \circ f_5\circ f_4^\ast \circ f_3\circ f_2^\ast \circ f_1 (\phi_C) =
   \langle \Phi_3\otimes \Phi_5 \otimes\Phi_7\otimes \Phi_8 |\phi_C \otimes
  \Phi_6 \otimes \Phi_4 \otimes \Phi_1\rangle,
  \nonumber\ena
and so the quantum information flow is encoded in the the form \eqa
  & & f_8^\ast\circ f_7\circ f_6^\ast \circ f_5\circ f_4^\ast
  \circ f_3\circ f_2^\ast \circ f_1 (\phi_C) \nonumber\\
  & &= \langle \Phi_3 \otimes \Phi_5 \otimes \Phi_7\otimes \Phi_8 |\phi_C \otimes
  \Phi_6 \otimes \Phi_4 \otimes \Phi_1 \otimes \Phi_2\rangle.
 \ena
 Namely, it is a composition of a series of maps,
\eq
 |\phi\rangle_B= f_8^\ast\circ f_7\circ f_6^\ast \circ f_5\circ f_4^\ast
  \circ f_3\circ f_2^\ast \circ f_1  |\phi\rangle_C
\en where the tensor product $|\Phi\rangle\otimes 1\!\! 1_d \otimes
\cdots \otimes 1\!\! 1_d$ is identified with $|\Phi\rangle$.
Additionally, following rules of the teleportation topology
 \cite{kl04a,kauffman05} to assign matrices $M, N$ to a cup
 and a cap respectively, we have the quantum information flow in the
  matrix formulation,
 \eq
 |\phi\rangle_B= M_8\cdot N_7\cdot M_6 \cdot N_5\cdot M_4
  \cdot N_3\cdot M_2 \cdot N_1  |\phi\rangle_C.
 \en

\begin{figure}
\begin{center}
\epsfxsize=12.cm \epsffile{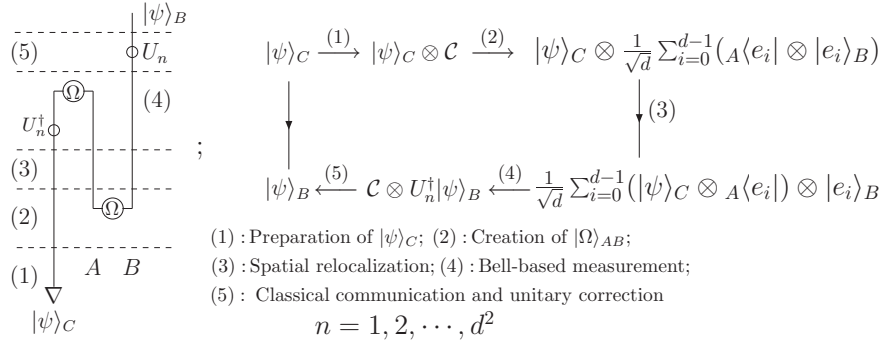} \caption{Quantum
information flow in strongly compact closed categories.}
\label{fig29}
\end{center}
\end{figure}

 The quantum information flow in terms of a composition of maps
 naturally leads to its description in the category theory. Here we
 show one-to-one correspondences between the quantum information
 flow and strongly compact categories.
 To transport Charlie's unknown quantum state $|\psi\rangle_C$ to Bob, the
  teleportation has to complete all the operations: the preparation of
  $|\psi\rangle_C$; the creation of $|\Omega\rangle_{AB}$ in Alice and
  Bob's system; the Bell-based measurement ${}_{CA}\langle \Omega_n|$ in
  Charlie and Alice's system; classical communications between Alice
  and Bob; Bob's local unitary corrections. These steps divide the
  quantum information flow into six pieces, and they are shown in the
  left diagrammatical term of Figure 29 where the third piece
  represents a process bringing Alice and Charlies' systems
  together for entangling measurement. In the category theory,
  every step (or piece) is denoted by a specific map satisfying
  the axioms of strongly compact closed categories. A
  crucial point is to recognize a bijective correspondence between
  a Bell state and a map from the dual Hilbert space ${\cal H}^\ast$
  to the Hilbert space $\cal H$,
  \eq
 \frac 1 {\sqrt d}\sum_{i=0}^{d-1}
 |e_i\rangle_{A}\otimes |e_i\rangle_{B}\approx
    \frac 1 {\sqrt d}\sum_{i=0}^{d-1}
 {}_A\langle e_i |\otimes |e_i\rangle_{B}, \qquad {\cal H}_A\otimes
 {\cal H}_B \approx {\cal H}_A^\ast \otimes {\cal H}_B
  \en
so that the  quantum information flow have a physical realization in
strongly compact closed categories. See the right
 term of Figure 29: the symbol
 ${\mathbb C}$ denotes the complex field and
 \eq
 |\psi\rangle_C\approx  |\psi\rangle_C \otimes {\mathcal C}, \qquad
 |\psi\rangle_B\approx  {\mathcal  C} \otimes |\psi\rangle_B
 \en
which suggests a bipartite state is created from vacuum denoted by a
complex number  and is annihilated it into the vacuum.

\subsection{Quantum information flow in the extended TL category}

\begin{figure}
\begin{center}
\epsfxsize=9.cm \epsffile{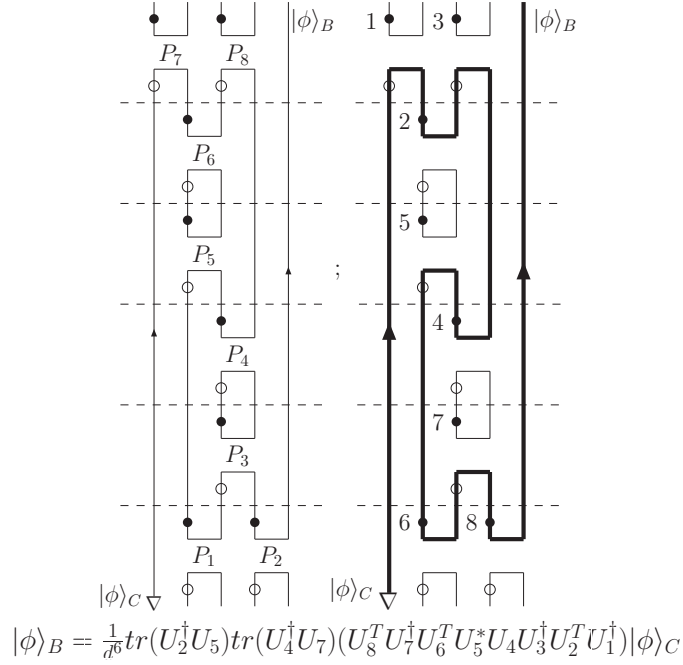} \caption{Quantum information
flow in the extended TL category.} \label{fig28}
\end{center}
\end{figure}

We revisit the example in Figure 28 and redraw a diagram, Figure 30,
according to the extended TL diagrammatical rules. Every projector
consists of a top cup and a bottom cap. Solid points $1,\cdots 8$ on
the left branches of cups respectively  denote local unitary
transformations $U_1, \cdots, U_8$, and small circles on the left
branches of caps denote their adjoint operators $U_1^\dag,\cdots,
U_8^\dag$, respectively. The quantum information flow from
$|\phi\rangle_C$ to $|\phi\rangle_B$ is determined by the transfer
operator,
 \eq
 \label{ourflow}
|\phi\rangle_B=\frac 1 {d^6} tr(U_2^\dag U_5) tr(U_4^\dag U_7)
 (U_8^T U_7^\dag  U_6^T  U_5^\ast  U_4  U_3^\dag
  U_2^T  U_1^\dag) |\phi\rangle_C
 \en
with the normalization factor $\frac 1 {d^6}$ contributed from six
vanishing cups and six vanishing caps as well as two traces from two
closed circles.

Five remarks are made as Figure 28 is compared with Figure 30. 1) In
the categorical approach, only the half of a projector is exploited
to use the bijective correspondence between a bipartite vector and a
map to represent the quantum information flow in terms of maps. In
the extended TL diagrammatical approach, however,  a projector is
denoted as the combination of a top cup and a bottom cap instead of
a single cup (or cap). Hence the quantum information flow from
$|\phi\rangle_C$ to $|\phi\rangle_B$ in Figure 30 has a
normalization factor contributed from closed circles which is
crucial for the quantum formation flow. For examples, setting eight
local unitary operators $U_i$ to be identity leads to $
|\phi\rangle_B =\frac 1 {d^4} |\phi\rangle_C $, and assuming $U_2$
and $U_5$ (or $U_4$ and $U_7$) orthogonal to each other causes a
zero vector to be sent to Bob, $|\phi\rangle_B=0$, no flow! 2) The
quantum information flow is only one part of the entire diagram in
the extended TL diagrammatical approach. Hence the acausality
problem of the quantum information flow in the categorical approach
is not reasonable since the whole process is not considered from the
global view. 3) In the extended TL diagrammatical approach, the
bijective correspondence between a local unitary transformation and
a bipartite vector is considered, which is different from the choice
in the categorical approach. For example, we have \eq
|\psi(U)\rangle=(U\otimes 1\!\! 1_d) |\Omega\rangle, \qquad
|\psi(U)\rangle \approx U \approx |\psi(U)\rangle \langle \psi(U)|,
\en and so Bell states (\ref{local}) are labeled by
 identity or Pauli matrices \eq
 |\phi^+\rangle  \approx 1\!\! 1_2, \,\,\, |\phi^-\rangle  \approx \sigma_3,
 \,\,\,
 |\psi^+\rangle  \approx \sigma_1,\,\,\, |\psi^-\rangle  \approx i
 \sigma_2.
\en As a projector is labeled by a local unitary transformation,
 the equation (\ref{ourflow}) is called the quantum information flow
 in terms of local unitary transformations (instead of maps). 4)
 The quantum information flow in the categorical approach  is created
 in view of additional permitted and forbidden rules \cite{coecke04},
 whereas it is  derived in a natural way without
imposed rules in the extended TL diagrammatical approach. 5) ${\cal
H}^\ast\otimes {\cal H}$ is imposed by the axioms of strongly
compact closed categories, see Figure 29, but is not required by
quantum teleportation,  the quantum information flow, and the
extended TL category.

 \section{Concluding remarks}

In this paper, we study algebraic structures and low dimensional
topology underlying quantum information and computation involving
maximally entangled states and local unitary transformations. We
describe quantum teleportation from the points of the symmetric
group, the braid group, the virtual braid group, the TL algebra and
the Brauer algebra, and propose the teleportation swapping, the
braid teleportation and the virtual braid teleportation. Especially,
quantum teleportation can be performed using the teleportation
swapping and Bell measurements, which is a description of quantum
teleportation via the Brauer algebra. Besides, we propose the
extended TL diagrammatical approach to study a series of topics: the
transfer operator with the acausality problem; measurement-based
quantum teleportation; tight teleportation and dense coding schemes;
the diagrammatical representation of the TL algebra; quantum
teleportation and the Brauer algebra; entanglement swapping;
universal quantum computing;  multipartite entanglements; and
quantum information flow. All these examples show that the extended
TL category is a mathematical framework describing quantum
information and computation using maximally entangled states and
local unitary transformations. For example, various descriptions to
quantum teleportation can be unified in the extended TL
diagrammatical approach.

As a further comment concluding this paper, we remark our previous
work \cite{zk07} on categorical foundation of quantum physics and
information. We suggest {\em unitary Hermitian ribbon categories} as
a natural extension of strongly compact closed categories with the
extended TL category as its special example. All known
diagrammatical approaches \cite{kl04a, kauffman05, coecke04, ac04,
coecke05,gwyc05, zhang06a, zhang06b,kl06, zk07} to quantum
information and computation including the extended TL diagrammatical
approach can be viewed as different versions of the diagrammatical
representation of tensor categories. Obviously, our future research
is focused to look for new quantum information protocols or quantum
algorithms with the help of mathematical structures presented in
this paper, especially the extended TL category.

 \section*{Acknowledgements}

 The author thanks L.H. Kauffman  and
 Y.-S. Wu for helpful comment.
 He is in part supported by the seed funding of
 University of Utah and NSFC Grant-10605035.


\begin{thebibliography}{99}

 \bibitem{werner89} R.F. Werner, {\it Quantum States with Einstein-Podolsky-Rosen
   Correlations Admitting a Hidden-Variable Model},
   Phys. Rev. A {\textbf 40} (1989) 4277.

  \bibitem{nielsen99}
  M. Nielsen and I. Chuang, {\it Quantum Computation and Quantum Information}
 (Cambridge University Press, 1999).

   \bibitem{mermin07} N.D. Mermin, {\it Quantum Computer Science}
  (Cambridge University Press, 2007).

 \bibitem{shor94} P.W. Shor, {\it  Algorithms for Quantum Computation:
  Discrete Logrithms and Factoring}. In S. Goldwasser, editor, {\em
  Proceedings of the 35th Annual Symposium on the Foundations of Computer
  Science}, pp. 124-134, Los Alamitos, CA, 1994. IEEE Computer
  Society Press.

  \bibitem{grover97} L.K. Grover, {\it Quantum Mechanics Helps in
   Searching for a Needle in a Haystack},
    Phys. Rev. Lett. {\bf 78} (1997) 325-328.

  \bibitem{bb84} C.H. Bennett, G. Brassard, {\it Quantum
  Cryptography: Public Key Distribution and Coin Tossing}, Int.
  Conf. Computers, Systems \& Signal Processing, Bangalore, India,
  December 10-12, 1984, pp. 175-179.

  \bibitem{ekert91} A.K. Ekert, {\it Quantum Cryptography Based on Bell's
  Theorem}, Phys. Rev. Lett. {\bf 67} (1991) 661-663.

  \bibitem{bbcjpw93} C.H. Bennett, G. Brassard, C. Crepeau, R.
  Jozsa, A. Peres and W. K. Wootters, {\it  Teleporting an Unknown Quantum State
  via Dual Classical and Einstein-Podolsky-Rosen Channels}, Phys. Rev. Lett. {\bf 70}
  (1993) 1895-1899.

  \bibitem{bdms00} S.L. Braunstein, G.M. D'Ariano, G.J. Milburn
 and M.F. Sacchi, {\it Universal Teleportation with a Twist}, Phys.
 Rev. Let. {\bf 84} (2000) 3486--3489.

  \bibitem {kauffman02} L.H. Kauffman, {\it Knots and Physics}
  (World Scientific Publishers, 2002).

  \bibitem{aravind97}
 P.K. Aravind, {\it Borromean Entanglement of the GHZ state}, in
 {\em Potentiality, Entanglement and Passion-at-a-Distance}, R.S. Cohen,
  M. Horne, and J. Stachel (eds.), pp. 53-59, Kluwer
 Academic Publishers, Boston 1997.

 \bibitem{dye02}
 H.A. Dye, {\it Unitary Solutions to the Yang--Baxter Equation in Dimension Four},
 Quant. Inf. Proc. {\bf 2} (2003) 117-150. Arxiv:  quant-ph/0211050.

 \bibitem{kl04a}
 L.H. Kauffman and S.J. Lomonaco Jr.,  {\it Braiding Operators are
 Universal Quantum Gates},  New J. Phys. {\bf 6} (2004) 134.
 Arxiv: quant-ph/0401090.

 \bibitem{bb02}
 J.L. Brylinski and R. Brylinski, {\it Universal quantum gates,} in {\em
Mathematics of Quantum Computation}, Chapman \& Hall/CRC Press, Boca
Raton, Florida, 2002 (edited by R. Brylinski and G. Chen).

 \bibitem{yang67} C.N. Yang,
 {\it Some Exact Results for the Many Body Problems in One Dimension with
 Repulsive Delta Function Interaction}, Phys. Rev. Lett. {\bf 19} (1967) 1312-1314.

 \bibitem{baxter72} R.J. Baxter, {\it Partition Function of the Eight-Vertex Lattice Model},
 Annals Phys.\ {\bf 70} (1972) 193-228.

 \bibitem{zkg05a} Y. Zhang, L.H. Kauffman and M.L. Ge,
 {\it Universal Quantum Gate, Yang--Baxterization and Hamiltonian}.
  Int. J. Quant. Inform., vol. 3, {\bf 4} (2005) 669-678. Arxiv: quant-ph/0412095.

 \bibitem{zkg05b} Y. Zhang, L.H. Kauffman and M.L. Ge,
 {\it Yang--Baxterizations, Universal Quantum Gates and
  Hamiltonians}.  Quant. Inf. Proc., vol. 4,{\bf 3} (2005) 159-197. Arxiv: quant-ph/0502015.

\bibitem{kauffman02a}
 L.H. Kauffman, {\it Quantum Computation and the Jones Polynomial}, in
 {\em Quantum Computation and Information}, S. Lomonaco, Jr. (ed.), AMS CONM/305, 2002,
  pp. 101-137. Arxiv: math. QA/0105255.

\bibitem {kauffman02b}
 L.H. Kauffman, {\it Quantum Topology and Quantum Computing}, in
{\em Quantum Computation}, S. Lomonaco (ed.), AMS PSAPM/58, 2002,
pp. 273--303.

 \bibitem{kl04b}
L. H. Kauffman and S. J. Lomonaco Jr., {\it Quantum Knots},
 in E. Donkor, A.R. Pirich and H.E. Brandt (eds.), Quantum Information
 and Computation II, Spie Proceedings, (12 -14 April, Orlando, FL, 2004),
 Vol. 5436, pp. 268-284. Arxiv: quant-ph/0403228.

\bibitem{kl04c}
 L.H. Kauffman and S.J. Lomonaco Jr., {\it Quantum Entanglement and
Topological Entanglement},  New J. Phys. {\bf 4} (2002) 73.1--73.18.

 \bibitem{kl03a}
L.H. Kauffman and S.J. Lomonaco Jr.,
 {\it Entanglement Criteria--Quantum and Topological}, in E. Donkor, A.R. Pirich
 and H.E. Brandt (eds.), Quantum Information and Computation -- Spie Proceedings,
 (21-22 April, Orlando, FL, 2003), Vol. 5105, pp. 51-58. Arxiv: quan-ph/0304091.

  \bibitem{zjg06} Y. Zhang, N. Jing and  M.L. Ge,
 {\it Quantum Algebras Associated with Bell States}.  J.Phys. A: Math.
 Theor. {\bf 41} (2008) 055310.

 \bibitem{frw06} J. Franko, E.C. Rowell and Z. Wang,
 {\it Extraspecial 2-Groups and Images of Braid Group
 Representations}. J. Knot Theory Ramifications, {\bf 15} (2006) 413-428.

  \bibitem{zg07} Y. Zhang and M.L. Ge,
{\it GHZ States, Almost-Complex Structure and Yang--Baxter
Equation}. Quant. Inf. Proc. vol. {\bf 6}, no. 5, (2007) 363-379.

 \bibitem{zrwwg07}  Y. Zhang, E.C. Rowell,  Y.-S Wu, Z. Wang and
  M.L. Ge, {\it From Extraspecial Two-Groups To GHZ States}.
  Arxiv: quant-ph/0706.1761.

  \bibitem{zhang08} Y. Zhang, {\it Quantum Error Correction Code in
  the Hamiltonian Formulation}. Arxiv: 0801.2561.

 \bibitem{kauffman05}
 L.H. Kauffman, {\it Teleportation Topology}.
 Opt. Spectrosc. {\bf 9} (2005) 227-232. Arxiv: quan-ph/0407224.

 \bibitem{werner01} R. F. Werner,  {\it All Teleportation and Dense Coding
  Schemes}, J. Phys. A {\bf 35} (2001) 7081--7094.  Arxiv: quant-ph/0003070.

 \bibitem{zkw05} Y. Zhang, L.H. Kauffman and R.F. Werner,
 {\it Permutation and its Partial Transpose}.
 Int. J. Quant. Inform.  vol. {\bf 5}, no. 4 (2006) 469-507.

\bibitem{zhang06a} Y. Zhang, {\it Teleportation, Braid Group and
   Temperley--Lieb Algebra}.  J.Phys. A:  Math. Theor.  {\bf 39} (2006)
   11599-11622.

 \bibitem{zhang06b} Y. Zhang, {\it Algebraic Structures Underlying
  Quantum Information Protocols}. ArXiv: quant-ph/0601050v2.


 \bibitem{zk07} Y. Zhang and L.H. Kauffman,
  {\it Topological-Like Features in Diagrammatical Quantum
  Circuits},  Quant. Inf. Proc. vol. {\bf 6}, no. 5  (2007) 477-507.

 \bibitem{tl71} H.N.V. Temperley and E.H. Lieb, {\it Relations between
  the `Percolation' and `Colouring' Problem and Other Graph-Theoretical Problems
  Associated with Regular Planar Lattices: Some Exact Results for the `Percolation'
  Problem}, Proc. Roy. Soc. A {\bf 322} (1971) 251-280.

  \bibitem{bell64} J.S. Bell, {\it On the Einstein-Podolsky-Rosen
  paradox}, Physics {\bf 1} (1964) 195-200.

 \bibitem{aav86} Y. Aharonov, D.Z. Albert, and L. Vaidman,
   {\it Measurement Process in Relativistic Quantum Theory},  Phys. Rev. {\bf D} 34
   (1986) 1805-1813.

 \bibitem{vaidman94} L. Vaidman, {\it Teleportation of Quantum States},
 Phys. Rev. {\bf A} 49 (1994) 1473-1475.

  \bibitem{vaidman03} L. Vaidman, {\it Instantaneous Measurement of Nonlocal
  Variables},  Phys. Rev. Lett. 90 (2003) 010402

   \bibitem{kauffman99} L.H. Kauffman,  {\it Virtual
 Knot Theory}, European J. Comb. {\bf 20} (1999) 663-690.

 \bibitem{brauer37}  R. Brauer, {\it On Algebras Which are Connected With the
 Semisimple Continuous Groups}, Ann. of Math. {\bf 38} (1937) 857-872.

 \bibitem{zkg06} Y. Zhang, L.H. Kauffman and M.L. Ge,
 {\it Virtual Extension of Temperley--Lieb Algebra}.
  ArXiv: math-ph/0610052

 \bibitem{preskill} J. Preskill, {\it Quantum Information and Computation},
 Lecture Notes for Ph219/CS219, Chapter 4, pp. 26-35.

 \bibitem{ekert93}
M. \.{Z}ukowski, A. Zeilinger, M.A. Horne and A.K. Ekert,  {\it
`Event-Ready-Detectors' Bell Experiment via Entanglement Swapping}.
Phys. Rev. Let. {\bf 71} (1993) 4287--4290.

\bibitem{gc99}
 D. Gottesman and I.L. Chuang,  {\it Quantum Teleportation is
a Universal Computational Primitive}.  Nature {\bf 402} (1999)
390--393. Arxiv:quant-ph/9908010.

 \bibitem{coecke04}  B. Coecke,  {\it The Logic of Entanglement. An
Invitation}. Oxford University Computing Laboratory Research Report
nr. PRG-RR-03-12. An 8 page short version is at
Arxiv:quant-ph/0402014.  The full 160 page version is at
\texttt{web.comlab.ox.ac.uk/oucl/ publications/tr/rr-03-12.html}.

 \bibitem{wilczek90} F. Wilczek, {\it Fractional Statistics and Anyon
 Superconductivity} (World Scientific, 1990).

 \bibitem{wu92} F.Y. Wu, {\it Knot Theory and Statistical Mechanics},
  Rev. Mod. Phys. {\bf 64} (1992) 1099-1131.

 \bibitem{jones87} V.F.R. Jones, {\it Heck Algebra Representations of Braid Groups
  and Link Polynomials},  Ann. of Math. {\bf 126} (1987) 335-388.

\bibitem{eriz05}  N. Erez, {\it Teleportation from a Projection Operator Point of
 View.}  Arxiv: quant-ph/0510130.

\bibitem{shor96}
 P.W. Shor, {\it Fault-Tolerant Quantum Computation}. In
 Proceedings, 35th Annual Symposium on Fundamentals of Computer
 Science (IEEE Press, Los Alamitos, 1996) 56-65. Arxiv:
 quant-ph/9605011.

 \bibitem{preskill97} J. Preskill, {\it Fault-Tolerant Quantum
 Computation}. Arxiv: quant-ph/9712048.

  \bibitem{aharonov05}   D. Aharonov, V. Jones and Z. Landau,
 {\it A Polynomial Quantum Algorithm for Approximating the Jones
 Polynomial}.   Arxiv: quant-ph/0511096.

 \bibitem{kitaev97} A. Yu. Kitaev, {\it Fault-Tolerant Quantum Computation by
 Anyons}, Annals Phys. {\bf 303} (2003) 2-30. Arxiv:
 quant-ph/9707021.

  \bibitem{flw02}  M.H. Freedman, M.J. Larsen and Z. Wang,
  {\it The Two-Eigenvalue Problem and Density of Jones Representation of Braid
  Groups},  Comm. Math. Phys. {\bf 228} (2002) 177-199.

  \bibitem{ac04} S. Abramsky, and B. Coecke,   {\it A Categorical
Semantics of Quantum Protocols}. In:  Proceedings of the 19th Annual
IEEE Symposium on Logic in Computer Science  (LiCS`04), IEEE
Computer Science Press. Arxiv:quant-ph/0402130.

  \bibitem{coecke05} B. Coecke,  {\it Kindergarten Quantum Mechanic--leture notes}.
 In: {\em Quantum Theory: Reconstructions of the Foundations III}, pp. 81-98,
 A. Khrennikov,  American Institute of Physics Press.
 Arxiv: quant-ph/0510032.

 \bibitem{gwyc05} R.B. Griffiths, S. Wu, L. Yu and S. M. Cohen,
  {\it Atemporal Diagrams for Quantum Circuits},
  Phys. Rev. {\bf A 73} (2006) 052309. Arxiv: quant-ph/0507215

  \bibitem{kl06} L.H. Kauffman and S.J. Lomonaco Jr.,
 {\it $q$-Deformed Spin Networks, Knot Polynomials and Anyonic
 Topological Quantum Computation}, Arxiv: quant-ph/0606114.



 \end{thebibliography}
\end{document}